\documentclass{cernyrep}

\def\boldvec#1{\ifmmode
\mathchoice{\mbox{\boldmath$\displaystyle\bf#1$}}
{\mbox{\boldmath$\textstyle\bf#1$}}
{\mbox{\boldmath$\scriptstyle\bf#1$}}
{\mbox{\boldmath$\scriptscriptstyle\bf#1$}}\else
{\mbox{\boldmath$\bf#1$}}\fi}

\begin{document}
\title{Topics in statistical data analysis for high-energy physics}
\author{G.~Cowan}
\institute{Royal Holloway, University of London, Egham, Surrey, TW20 0EX, UK}
\maketitle

\begin{abstract}
These lectures concern two topics that are becoming increasingly
important in the analysis of High Energy Physics (HEP) data: Bayesian
statistics and multivariate methods.  In the Bayesian approach we
extend the interpretation of probability to cover not only the
frequency of repeatable outcomes but also to include a degree of
belief.  In this way we are able to associate probability with a
hypothesis and thus to answer directly questions that cannot be
addressed easily with traditional frequentist methods.  In
multivariate analysis we try to exploit as much information as
possible from the characteristics that we measure for each event to
distinguish between event types.  In particular we will look at a
method that has gained popularity in HEP in recent years: the boosted
decision tree (BDT).
\end{abstract}

\section{Introduction}
\label{sec:intro}

When a high-energy physics experiment enters the phase of data collection and
analysis, the daily tasks of its postgraduate students are often
centred not around the particle physics theories one is trying to test
but rather on statistical methods.  These methods are the tools needed
to compare data with theory and quantify the extent to which one
stands in agreement with the other.  Of course one must understand the
physical basis of the models being tested and so the theoretical
emphasis in postgraduate education is no doubt well founded.  But with
the increasing cost of HEP experiments it has become important to
exploit as much of the information as possible in the hard-won data,
and to quantify as accurately as possible the inferences one draws
when confronting the data with model predictions.

Despite efforts to make the lectures self contained, some familiarity
with basic ideas of statistical data analysis is assumed.
Introductions to the subject can be found, for example, in the reviews
of the Particle Data Group~\cite{PDG} or in the texts~\cite{Cowan98,Lyons,Barlow,James,Brandt}.

In these two lectures we will discuss two topics that are becoming
increasingly important: Bayesian statistics and multivariate methods.
In Section~\ref{sec:bayes} we will review briefly the concept of
probability and see how this is used differently in the frequentist
and Bayesian approaches.  Then in Section~\ref{sec:linefit} we will
discuss a simple example, the fitting of a straight line to a set of
measurements, in both the frequentist and Bayesian approaches and
compare different aspects of the two.  This will include in
Section~\ref{sec:MCMC} a brief description of Markov Chain Monte Carlo
(MCMC), one of the most important tools in Bayesian computation.  We
generalize the treatment in Section~\ref{sec:fitsys} to include
systematic errors.

In Section~\ref{sec:multivariate} we take up the general problem of how
to distinguish between two classes of events, say, signal and
background, on the basis of a set of characteristics measured for each
event.  We first describe how to quantify the performance of a
classification method in the framework of a statistical test.
Although the Neyman--Pearson lemma indicates that this problem has an
optimal solution using the likelihood ratio, this usually cannot be
used in practice and one is forced to seek other methods.  In
Section~\ref{sec:bdt} we look at a specific example of such a method, the
boosted decision tree.  Using this example we describe several issues
common to many classification methods, such as overtraining.  Finally,
some conclusions are mentioned in Section~\ref{sec:summary}.

\section{Bayesian statistical methods for high-energy physics}
\label{sec:bayes}

In this section we look at the basic ideas of Bayesian statistics and
explore how these can be applied in particle physics.  We will
contrast these with the corresponding notions in frequentist
statistics, and to make the treatment largely self contained, the main
ideas of the frequentist approach will be summarized as well.

\subsection{The role of probability in data analysis}
\label{sec:prob}

We begin by defining probability with the axioms written down by
Kolmogorov~\cite{Kolmogorov33} using the language of set theory.
Consider a set $S$ containing subsets $A, B, \ldots$.  We
define the probability $P$ as a real-valued function with the
following properties:

\begin{enumerate}
\item For every subset $A$ in $S$, $P(A) \ge 0$;
\item For disjoint subsets (i.e., $A \cap B = \emptyset$), 
$P(A \cup B) = P(A) + P(B)$;
\item $P(S) = 1$.
\end{enumerate}

\noindent In addition, we define the conditional probability $P(A|B)$ 
(read $P$ of $A$ given $B$) as 

\begin{equation}
\label{eq:condprobdef}
P(A|B) = \frac{ P(A \cap B)}{P(B)} \;.
\end{equation}

\noindent From this definition and using the fact that $A \cap B$ and
 $B \cap A$ are the same, we obtain {\it Bayes' theorem},

\begin{equation}
\label{eq:bayesthm}
P(A|B) = \frac{P(B|A) P(A)}{P(B)} \;.
\end{equation}

\noindent From the three axioms of probability and the definition of
conditional probability, we can derive the {\it law of total
probability},

\begin{equation}
\label{eq:totalprob}
P(B) = \sum_i P(B|A_i) P(A_i) \;,
\end{equation}

\noindent for any subset $B$ and for disjoint $A_i$ with 
$\cup_i A_i = S$.  This can be combined with Bayes' theorem~(\ref{eq:bayesthm}) to give

\begin{equation}
\label{eq:bayesthm2}
P(A|B) = \frac{P(B|A) P(A)}{\sum_i P (B|A_i) P(A_i)} \;,
\end{equation}

\noindent where the subset $A$ could, for example, be one of the $A_i$.

The most commonly used interpretation of the subsets of the sample
space are outcomes of a repeatable experiment.  The probability $P(A)$
is assigned a value equal to the limiting frequency of occurrence of
$A$.  This interpretation forms the basis of {\it frequentist
statistics}.

The subsets of the sample space can also be interpreted as {\it
hypotheses}, i.e., statements that are either true or false, such as
``The mass of the $W$ boson lies between 80.3 and 80.5 GeV.''  In the
frequency interpretation, such statements are either always or never
true, i.e., the corresponding probabilities would be 0 or 1.  Using
{\it subjective probability}, however, $P(A)$ is interpreted as the
degree of belief that the hypothesis $A$ is true.

Subjective probability is used in {\it Bayesian} (as opposed to
frequentist) statistics.  Bayes' theorem can be written

\begin{equation}
\label{eq:bayesthm3}
P(\hbox{theory}|\hbox{data}) \propto 
P(\hbox{data}|\hbox{theory}) P(\hbox{theory}) \;,
\end{equation}

\noindent where `theory' represents some hypothesis and `data' is the
outcome of the experiment.  Here $P(\hbox{theory})$ is the {\it prior}
probability for the theory, which reflects the experimenter's degree
of belief before carrying out the measurement, and
$P(\hbox{data}|\hbox{theory})$ is the probability to have gotten the
data actually obtained, given the theory, which is also called the
{\it likelihood}.

Bayesian statistics provides no fundamental rule for obtaining the
prior probability; this is necessarily subjective and may depend on
previous measurements, theoretical prejudices, \etc{} Once this has
been specified, however, Eq.~(\ref{eq:bayesthm3}) tells how the
probability for the theory must be modified in the light of the new
data to give the {\it posterior} probability,
$P(\hbox{theory}|\hbox{data})$.  As Eq.~(\ref{eq:bayesthm3}) is stated
as a proportionality, the probability must be normalized by summing
(or integrating) over all possible hypotheses.

The difficult and subjective nature of encoding personal knowledge
into priors has led to what is called {\it objective Bayesian
statistics}, where prior probabilities are based not on an actual
degree of belief but rather derived from formal rules.  These give,
for example, priors which are invariant under a transformation of
parameters or which result in a maximum gain in information for a
given set of measurements.  For an extensive review see, for example,
Ref.~\cite{KassWasserman96}.

\subsection{An example:  fitting a straight line}
\label{sec:linefit}

In Section~\ref{sec:linefit} we look at the example of a simple fit in
both the frequentist and Bayesian frameworks.  Suppose we have
independent data values $y_i$, $i=1,...,n$, that are each made at a
given value $x_i$ of a control variable $x$.  Suppose we model the
$y_i$ as following a Gaussian distribution with given standard
deviations $\sigma_i$ and mean values $\mu_i$ given by a function that
we evaluate at the corresponding $x_i$,

\begin{equation}
\label{eq:muofx}
\mu(x; \theta_0, \theta_1) = \theta_0 + \theta_1 x \;.
\end{equation}

\noindent We would like to determine values of the parameters
$\theta_0$ and $\theta_1$ such that the model best describes the data.
The ingredients of the analysis are illustrated in
Fig.~\ref{fig:linefit}(a).


\setlength{\unitlength}{1.0 cm}
\renewcommand{\baselinestretch}{0.9}
\begin{figure}[htbp]
\begin{picture}(10.0,6.5)
\put(0.5,0){\includegraphics{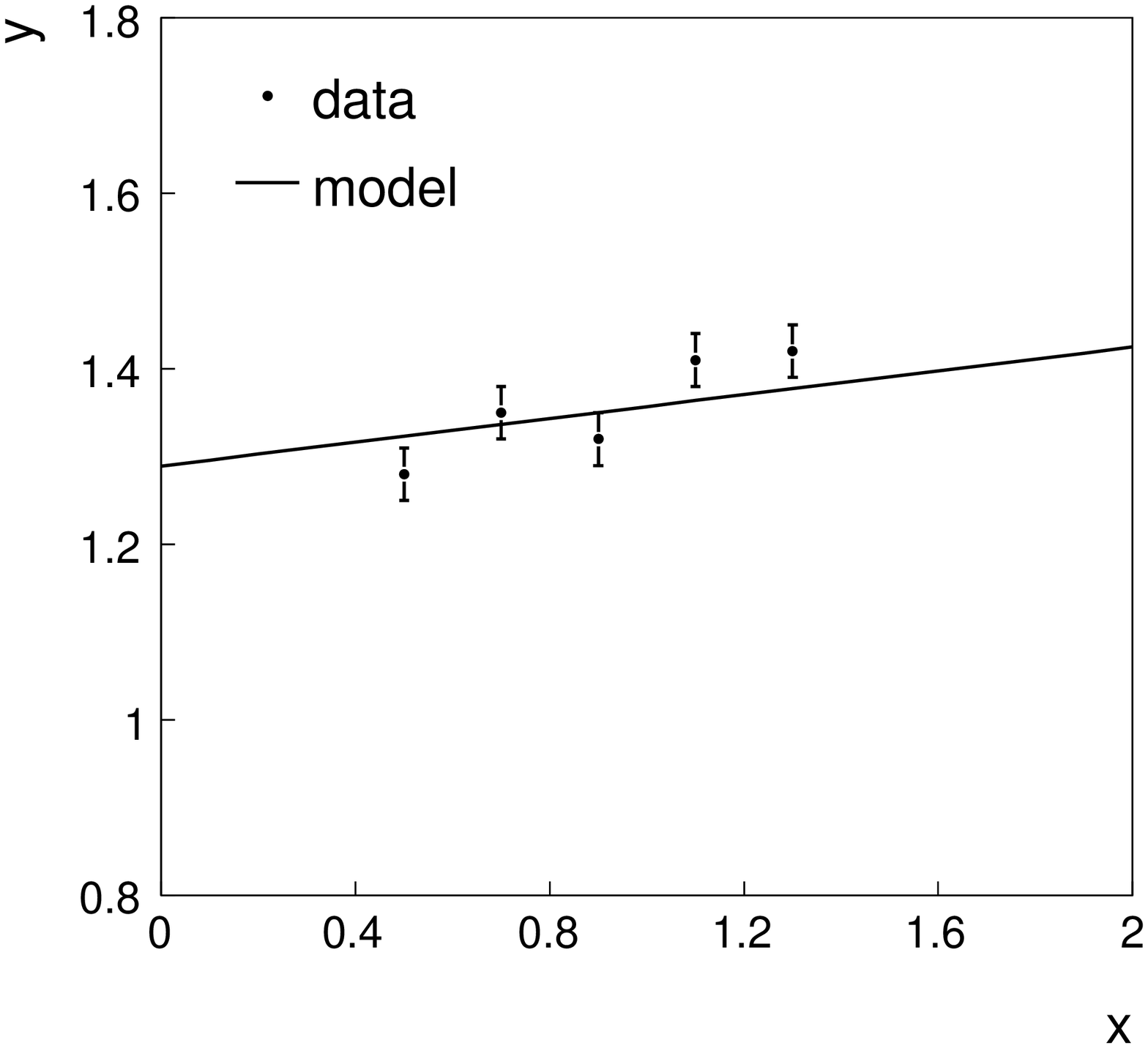}}
\put(8.5,0){\includegraphics{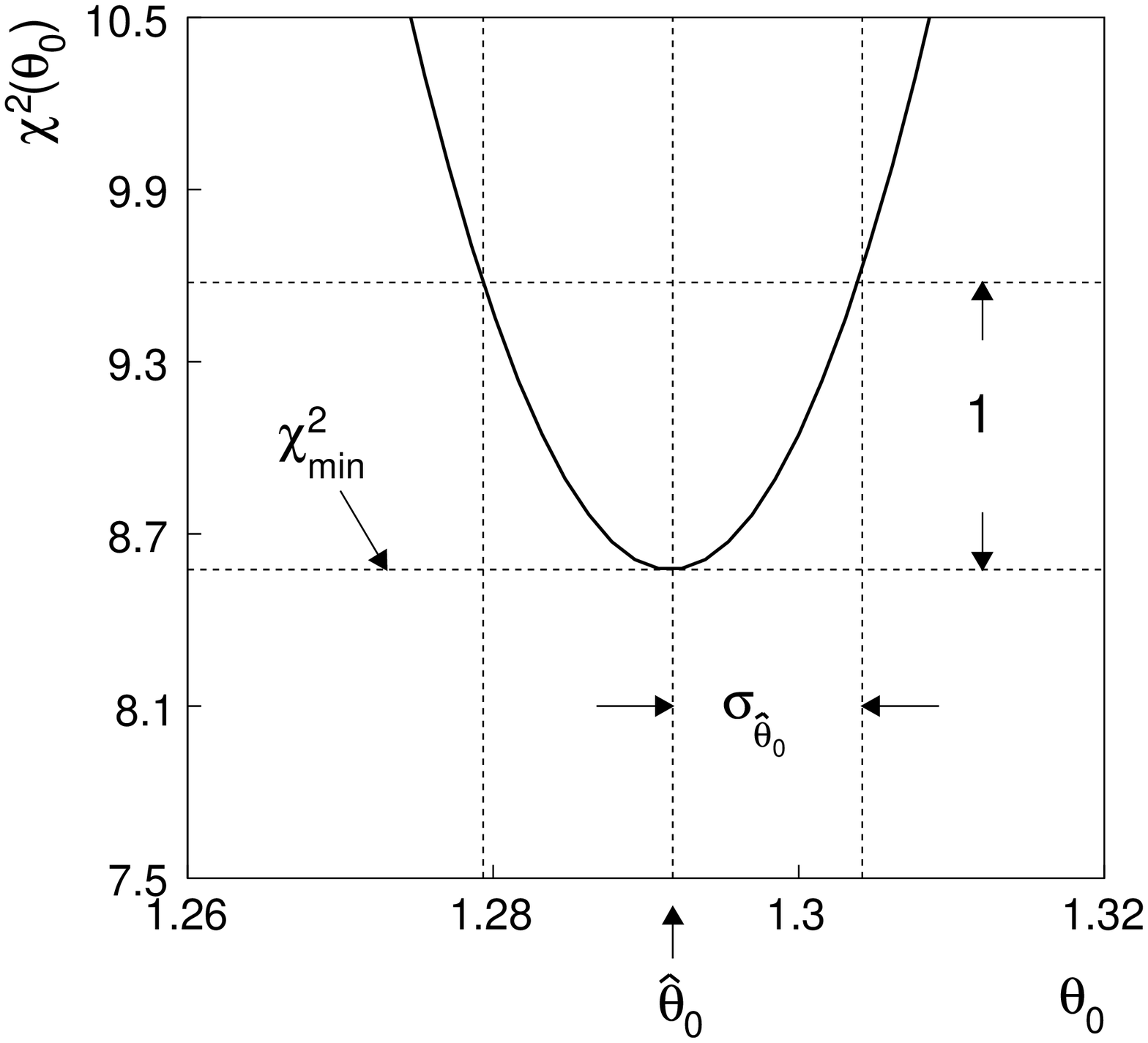}}
\put(0.,5.5){(a)}
\put(15.5,5.5){(b)}
\end{picture}
\caption{\small (a) Illustration of fitting a straight line 
to data (see text).
(b) The $\chi^2$ as a function of the parameter $\theta_0$, illustrating
the method to determine the estimator $\hat{\theta}_0$ and its
standard deviation $\sigma_{\hat{\theta}_0}$.}
\label{fig:linefit}
\end{figure}
\renewcommand{\baselinestretch}{1}
\small\normalsize

Now suppose the real goal of the analysis is only to estimate the
parameter $\theta_0$.  The slope parameter $\theta_1$ must also be
included in the model to obtain a good description of the data, but we
are not interested in its value as such.  We refer to $\theta_0$ as
the parameter of interest, and $\theta_1$ as a {\it nuisance
parameter}.  In the following sections we treat this problem using
both the frequentist and Bayesian approaches.

\subsubsection{The frequentist approach}
\label{sec:freqfit}

Our model states that the measurements are Gaussian distributed, i.e.,
the probability density function (pdf) for the $i$th measurement $y_i$
is

\begin{equation}
\label{eq:yipdf}
f(y_i; \boldvec{\theta}) = \frac{1}{\sqrt{2\pi} \sigma_i} 
e^{-(y_i - \mu(x_i; \boldvec{\theta}))^2 / 2 \sigma_i^2 } \;,
\end{equation}

\noindent where $\boldvec{\theta} = (\theta_0, \theta_1)$.  

The {\it likelihood function} is the joint pdf for all of the $y_i$,
evaluated with the $y_i$ obtained and regarded as a function of the
parameters.  Since we are assuming that the measurements are
independent, the likelihood function is in this case given by the
product

\begin{equation}
\label{eq:yilikelihood}
L(\boldvec{\theta}) = \prod_{i=1}^n f(y_i; \boldvec{\theta}) = 
\prod_{i=1}^n\frac{1}{\sqrt{2\pi} \sigma_i} 
e^{-(y_i - \mu(x_i; \boldvec{\theta}))^2 / 2 \sigma_i^2 }
\;.
\end{equation}

\noindent In the frequentist approach we construct estimators
$\hat{\boldvec{\theta}}$ for the parameters $\boldvec{\theta}$,
usually by finding the values that maximize the likelihood function.
(We will write estimators for parameters with hats.)  In this case one
can see from~(\ref{eq:yilikelihood}) that this is equivalent to
minimizing the quantity

\begin{equation}
\label{eq:chi2}
\chi^2(\boldvec{\theta}) = \sum_{i=1}^n
\frac{ (y_i -  \mu(x_i; \boldvec{\theta}))^2}{\sigma_i^2 } 
= - 2 \ln L(\boldvec{\theta}) + C \;,
\end{equation}

\noindent where $C$ represents terms that do not depend on the
parameters.  Thus for the case of independent Gaussian measurements,
the maximum likelihood (ML) estimators for the parameters coincide
with those of the method of least squares (LS).

Suppose first that the slope parameter $\theta_1$ is known exactly,
and so it is not adjusted to maximize the likelihood (or minimize the
$\chi^2$) but rather held fixed.  The quantity $\chi^2$ versus the
single adjustable parameter $\theta_0$ would be as shown in
Fig.~\ref{fig:linefit}(b), where the minimum indicates the value of
the estimator $\hat{\theta}_0$.

Methods for obtaining the standard deviations of estimators --- the
statistical errors of our measured values --- are described in many
references such as~\cite{PDG,Cowan98,Lyons,Barlow,James,Brandt}.  Here
in the case of a single fitted parameter the rule boils down to moving
the parameter away from the estimate until $\chi^2$ increases by one
unit (i.e., $\ln L$ decreases from its maximum by $1/2$) as indicated
in the figure.

It may be, however, that we do not know the value of the slope
parameter $\theta_1$, and so even though we do not care about its
value in the final result, we are required to treat it as an
adjustable parameter in the fit.  Minimizing
$\chi^2(\boldvec{\theta})$ results in the estimators
$\hat{\boldvec{\theta}} = (\hat{\theta}_0, \hat{\theta}_1)$, as
indicated schematically in Fig.~\ref{fig:Chi2Contour}(a).  Now the
recipe to obtain the statistical errors, however, is not simply a
matter of moving the parameter away from its estimated value until the
$\chi^2$ goes up by one unit.  Here the standard deviations must be
found from the tangent lines (or in higher-dimensional problems, the
tangent hyperplanes) to the contour defined by
$\chi^2(\boldvec{\theta}) = \chi^2_{\rm min} + 1$, as shown in the
figure.

\setlength{\unitlength}{1.0 cm}
\renewcommand{\baselinestretch}{0.9}
\begin{figure}[htbp]
\begin{picture}(10.0,6.5)
\put(0.5,0){\includegraphics{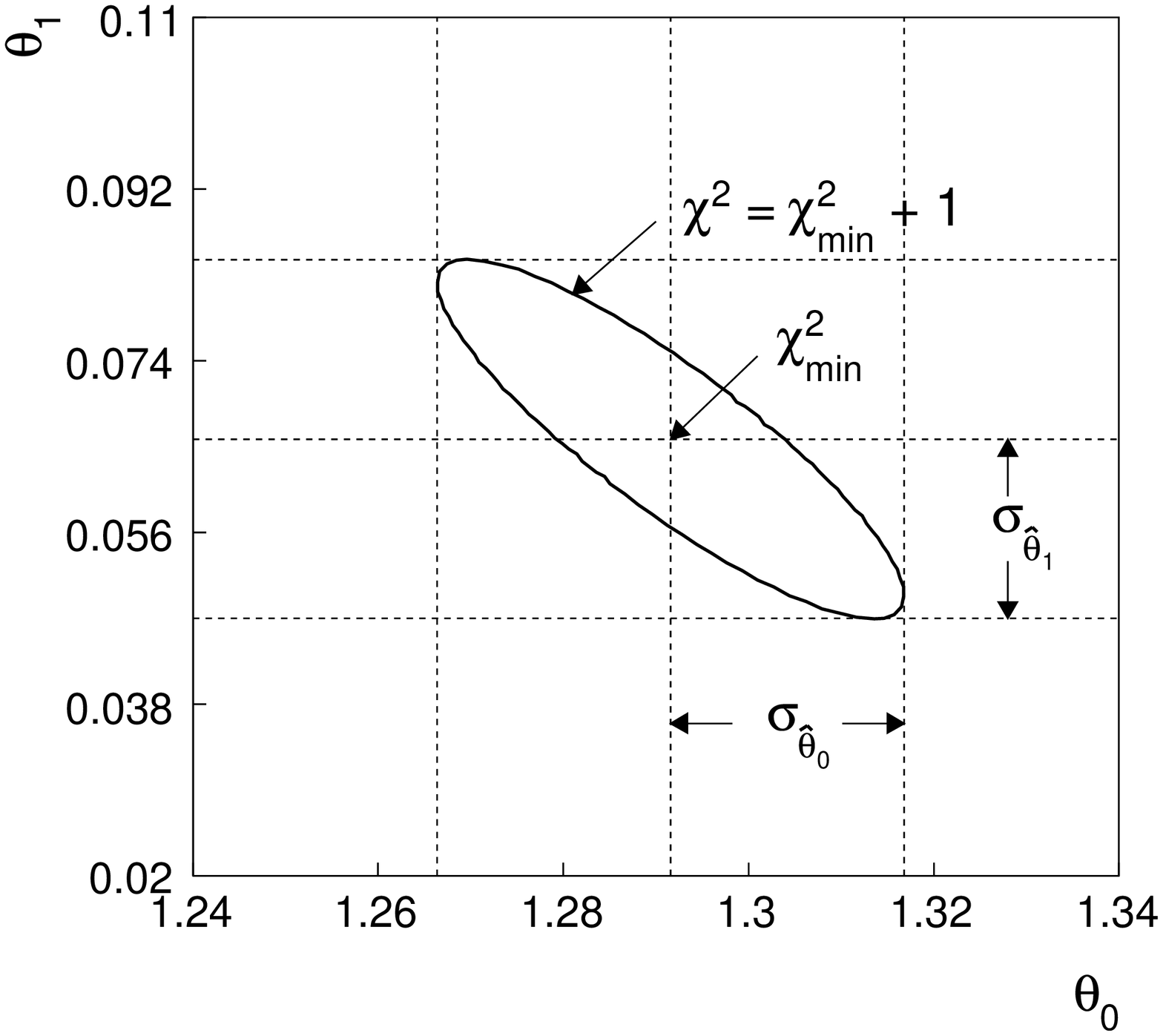}}
\put(8.5,0){\includegraphics{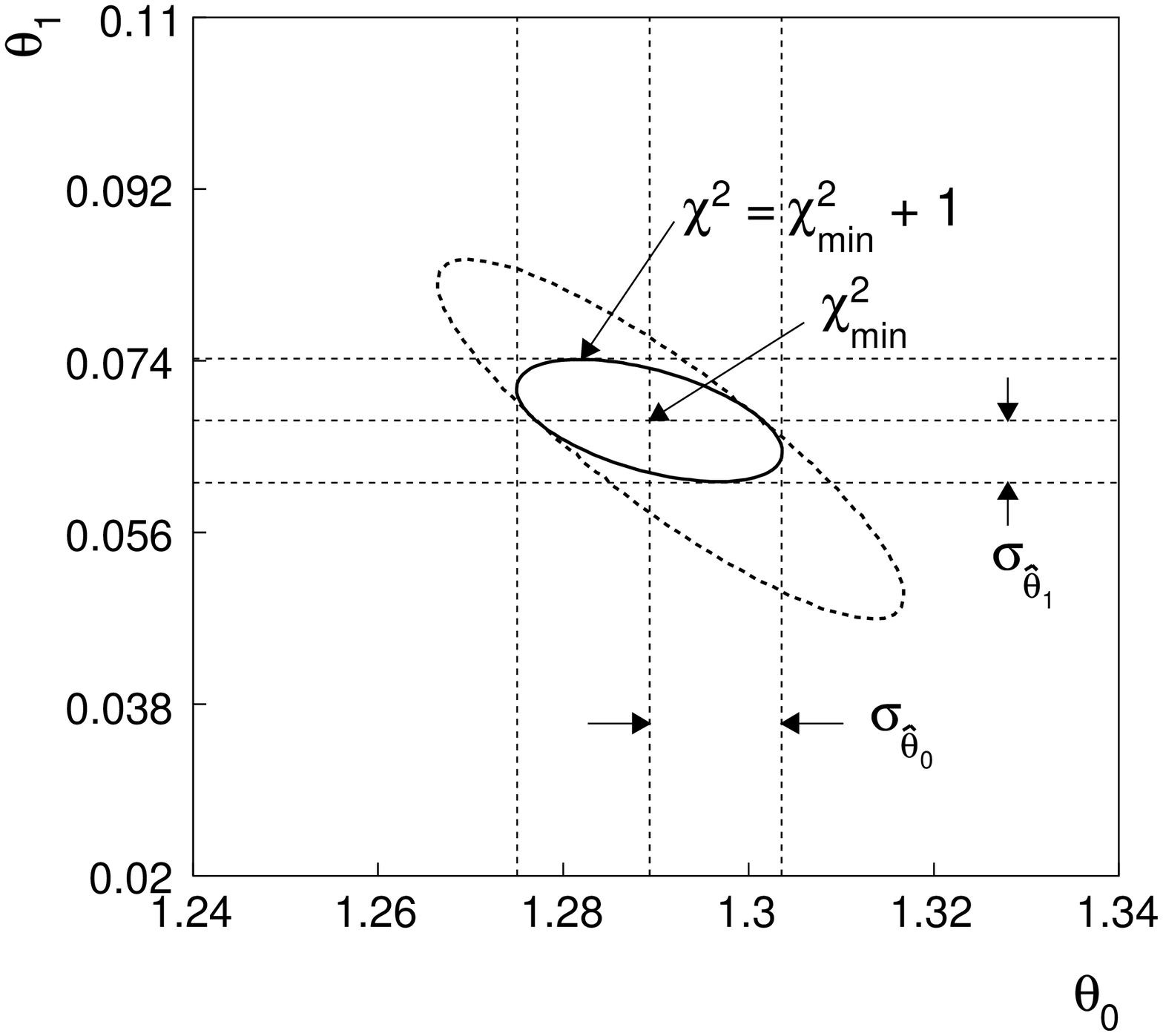}}
\put(0.,5.5){(a)}
\put(15.5,5.5){(b)}
\end{picture}
\caption{\small Contour of $\chi^2(\boldvec{\theta}) = 
\chi^2_{\rm min} + 1$
centred about the estimates $(\hat{\theta}_0, \hat{\theta}_1)$ (a)
with no prior measurement of $\theta_1$ and (b) when a prior 
measurement of $\theta_1$ is included.}
\label{fig:Chi2Contour}
\end{figure}
\renewcommand{\baselinestretch}{1}
\small\normalsize

The tilt of the contour in Fig.~\ref{fig:Chi2Contour}(a) reflects 
the correlation between the estimators $\hat{\theta}_0$ and
$\hat{\theta}_1$.  A useful estimate for the inverse of the matrix of
covariances $V_{ij} = \mbox{cov}[V_i, V_j]$ can be found from the
second derivative of the log-likelihood evaluated at its maximum,

\begin{equation}
\label{eq:covmat}
\widehat{V}^{-1}_{ij} = - \left. \frac{ \partial^2 \ln L }
{\partial \theta_i \partial \theta_j} \right|_{{\boldvec{\theta}} = 
\hat{\boldvec{\theta}}} \;.
\end{equation}

\noindent More information on how to extract the full covariance
matrix from the contour can be found, for example, in
Refs.~\cite{PDG,Cowan98,Lyons,Barlow,James,Brandt}.  The point to note
here is that the correlation between the estimators for the parameter
of interest and the nuisance parameter has the result of inflating the
standard deviations of both.  That is, if $\theta_1$ were known
exactly, then the distance one would have to move $\theta_0$ away from
its estimated value to make the $\chi^2$ increase by one unit would be
less, as one can see from the figure.  So although we can improve the
ability of a model to describe the data by including additional
nuisance parameters, this comes at the price of increasing the
statistical errors.  This is an important theme which we will
encounter often in data analysis.

Now consider the case where we have a prior measurement of $\theta_1$.
For example, we could have a measurement $t_1$ which we model as
following a Gaussian distribution centred about $\theta_1$ and having
a given standard deviation $\sigma_{t_{1}}$.  If this measurement is
independent of the other $y_i$ values, then the full likelihood
function is obtained simply by multiplying the original one by a
Gaussian, and so when we find the new $\chi^2$ from $-2 \ln L$ there
is an additional term, namely,

\begin{equation}
\label{eq:chi2Prev}
\chi^2(\boldvec{\theta}) = \sum_{i=1}^n
\frac{ (y_i -  \mu(x_i; \boldvec{\theta}))^2}{\sigma_i^2 } 
+ \frac{(\theta_1 - t_1)^2}{\sigma_{t_{1}}^2} \;.
\end{equation}

As shown in Fig.~\ref{fig:Chi2Contour}(b), the new (solid) contour of
$\chi^2 = \chi^2_{\rm min} + 1$ is compressed relative to the old
(dashed) one in the $\theta_1$ direction, and this compression has the
effect of decreasing the error in $\theta_0$ as well.  The lesson is:
by better constraining nuisance parameters, one improves the
statistical accuracy of the parameters of interest.

\subsubsection{The Bayesian approach}
\label{sec:bayesfit}

To treat the example above in the Bayesian framework, we write
Bayes' theorem~(\ref{eq:bayesthm}) as

\begin{equation}
\label{eq:bayesthmlinefit}
p(\boldvec{\theta}|\boldvec{y}) = \frac{ L(\boldvec{y} | \boldvec{\theta}) 
\pi(\boldvec{\theta}) }
{ \int L(\boldvec{y} | \boldvec{\theta}) \pi(\boldvec{\theta}) \, 
d \boldvec{\theta} } 
\;.
\end{equation}

\noindent Here $\boldvec{\theta} = (\theta_0, \theta_1)$ symbolizes
the hypothesis whose probability we want to determine.  The likelihood
$L(\boldvec{y} | \boldvec{\theta})$ is the probability to obtain the
data $\boldvec{y} = (y_1, \ldots, y_n)$ given the hypothesis, and the
prior probability $\pi(\boldvec{\theta}|\boldvec{y})$ represents our
degree of belief about the parameters before seeing the outcome of the
experiment.  The posterior probability $p(\boldvec{\theta})$
encapsulates all of our knowledge about $\boldvec{\theta}$ when the
data $\boldvec{y}$ is combined with our prior beliefs.  The
denominator in~(\ref{eq:bayesthmlinefit}) serves to normalize the
posterior pdf to unit area.

The likelihood $L(\boldvec{y} | \boldvec{\theta})$ is the same as the
$L(\boldvec{\theta})$ that we used in the frequentist approach above.  The
slightly different notation here simply emphasizes its role as the
conditional probability for the data given the parameter.

To proceed we need to write down a prior probability density
$\pi(\theta_0, \theta_1)$.  This phase of a Bayesian analysis,
sometimes called the {\it elicitation of expert opinion}, is in many
ways the most problematic, as there are no universally accepted rules
to follow.  Here we will explore some of the important issues that
come up.

In general, prior knowledge about one parameter might affect knowledge
about the other, and if so this must be built into $\pi(\theta_0,
\theta_1)$.  Often, however, one may regard the prior knowledge about
the parameters as independent, in which case the density factorizes as

\begin{equation}
\label{eq:pit0t1}
\pi(\theta_0, \theta_1) = \pi_0 (\theta_0) \pi_1(\theta_1) \;.
\end{equation}

\noindent For purposes of the present example we will assume that
this holds.  

For the parameter of interest $\theta_0$, it may be that we have
essentially no prior information, so the density $\pi_0 (\theta_0)$
should be very broad.  Often one takes the limiting case of a broad
distribution simply to be a constant, i.e.,

\begin{equation}
\label{eq:pit0}
\pi_0(\theta_0) = \mbox{const.} \;.
\end{equation}

\noindent Now one apparent problem with Eq.~(\ref{eq:pit0}) is that it
is not normalizable to unit area, and so does not appear to be a valid
probability density.  It is said to be an {\it improper prior}.  The
prior always appears in Bayes' theorem multiplied by the likelihood,
however, and as long as this falls off quickly enough as a function of
the parameters, then the resulting posterior probability density can
be normalized to unit area.

A further problem with uniform priors is that if the prior pdf is
flat in $\boldvec{\theta}$, then it is not flat for a nonlinear
function of $\boldvec{\theta}$, and so a different parametrization of
the problem would lead in general to a non-equivalent posterior pdf.

For the special case of a constant prior, one can see from Bayes'
theorem~(\ref{eq:bayesthmlinefit}) that the posterior is proportional
to the likelihood, and therefore the mode (peak position) of the
posterior is equal to the ML estimator.  The posterior mode, however,
will change in general upon a transformation of parameter.  A summary
statistic other than the mode may be used as the Bayesian estimator,
such as the median, which is invariant under a monotonic parameter
transformation.  But this will not in general coincide with the ML
estimator.

For the prior $\pi_1(\theta_1)$, let us assume that our prior
knowledge about this parameter includes the earlier measurement $t_1$,
which we modelled as a Gaussian distributed variable centred about
$\theta_1$ with standard deviation $\sigma_{t_{1}}$.  If we had taken,
even prior to that measurement, a constant prior for $\theta_1$, then
the ``intermediate-state'' prior that we have before looking at the
$y_i$ is simply this flat prior times the Gaussian likelihood, i.e., a
Gaussian prior in $\theta_1$:

\begin{equation}
\label{eq:pit1}
\pi_1(\theta_1) = \frac{1}{\sqrt{2 \pi} \sigma_{t_{1}}} 
e^{-(\theta_1 - t_1)^2/2\sigma_{t_{1}}^2} \;.
\end{equation}

Putting all of these ingredients into Bayes' theorem gives

\begin{equation}
\label{eq:bthmlf}
p(\theta_0, \theta_1 | \boldvec{y}) \propto
\prod_{i=1}^n \frac{1}{\sqrt{2 \pi} \sigma_i}
e^{-(y_i - \mu(x_i; \theta_0, \theta_1))^2 / 2 \sigma_i^2 } \,
\pi_0 \, \frac{1}{\sqrt{2 \pi} \sigma_{t_{1}}} 
e^{-(\theta_1 - t_1)^2/2\sigma_{t_{1}}^2} \;,
\end{equation}

\noindent where $\pi_0$ represents the constant prior in $\theta_0$
and the equation has been written as a proportionality with the
understanding that the final posterior pdf should be normalized to
unit area.

What Bayes' theorem gives us is the full joint pdf $p(\theta_0,
\theta_1 | \boldvec{y})$ for both the parameter of interest $\theta_0$
as well as the nuisance parameter $\theta_1$.  To find the pdf for the
parameter of interest only, we simply integrate (marginalize) the
joint pdf, i.e.,

\begin{equation}
\label{eq:marginal}
p(\theta_0 | \boldvec{y} ) = 
\int p (\theta_0,\theta_1 | \boldvec{y} ) \, d \theta_1 \;.
\end{equation}

\noindent In this example, it turns out that we can do the integral
in closed form.  We find a Gaussian posterior,

\begin{equation}
\label{eq:posterior}
p(\theta_0 | \boldvec{y}) = \frac{1}{\sqrt{2 \pi} \sigma_{\theta_{0}}}
e^{-(\theta_0 - \hat{\theta}_0)^2 / 2 \sigma_{\theta_{0}}^2} \;,
\end{equation}

\noindent where $\hat{\theta}_0$ is in fact the same as the ML (or LS)
estimator found above with the frequentist approach, and
$\sigma_{\theta_{0}}$ is the same as the standard deviation of that
estimator $\sigma_{\hat{\theta}_0}$.

So we find something that looks just like the frequentist answer,
although here the interpretation of the result is different.  The
posterior pdf $p(\theta_0 | \boldvec{y})$ gives our degree of belief
about the location of the parameter in the light of the data.  We will
see below how the Bayesian approach can, however, lead to results that
differ both in interpretation as well as in numerical value from what
would be obtained in a frequentist calculation.  First, however, we
need to pause for a short digression on Bayesian computation.

\subsubsection{Bayesian computation and MCMC}
\label{sec:MCMC}

In most real Bayesian calculations, the marginalization integrals
cannot be carried out in closed form, and if the number of nuisance
parameters is too large then they can also be difficult to compute
with standard Monte Carlo methods.  However, {\it Markov Chain Monte
Carlo} (MCMC) has become the most important tool for computing
integrals of this type and has revolutionized Bayesian computation.
In-depth treatments of MCMC can be found, for example, in the texts by Robert
and Casella~\cite{Robert04}, Liu~\cite{Liu01}, and the review by Neal~\cite{Neal93}.

The basic idea behind using MCMC to marginalize the joint pdf
$p(\theta_0, \theta_1|\boldvec{y})$ is to sample points
$\boldvec{\theta} = (\theta_0, \theta_0)$ according to the posterior
pdf but then only to look at the distribution of the component of
interest, $\theta_0$.  A simple and widely applicable MCMC method is
the Metropolis-Hastings algorithm, which allows one to generate
multidimensional points $\boldvec{\theta}$ distributed according to a
target pdf that is proportional to a given function
$p(\boldvec{\theta})$, which here will represent our posterior pdf.
It is not necessary to have $p(\boldvec{\theta})$ normalized to unit
area, which is useful in Bayesian statistics, as posterior probability
densities are often determined only up to an unknown normalization
constant, as is the case in our example.

To generate points that follow $p(\boldvec{\theta})$, one first needs
a proposal pdf $q(\boldvec{\theta}; \boldvec{\theta}_0)$, which
can be (almost) any pdf from which independent random values
$\boldvec{\theta}$ can be generated, and which contains as a parameter
another point in the same space $\boldvec{\theta}_0$.  For example, a
multivariate Gaussian centred about $\boldvec{\theta}_0$ can be used.
Beginning at an arbitrary starting point $\boldvec{\theta}_0$, the
Hastings algorithm iterates the following steps:

\begin{enumerate}

\item  Generate a value $\boldvec{\theta}$ using the proposal density 
$q(\boldvec{\theta}; \boldvec{\theta}_0)$;

\item Form the Hastings test ratio, $\alpha = \min \left[ 1, \frac{
p(\boldvec{\theta}) q(\boldvec{\theta}_0; \boldvec{\theta}) } {
p(\boldvec{\theta}_0) q(\boldvec{\theta}; \boldvec{\theta}_0) }
\right]$;

\item Generate a value $u$ uniformly distributed in $[0,1]$;

\item If $u \le \alpha$, take $\boldvec{\theta}_1 = \boldvec{\theta}$.
Otherwise, repeat the old point, i.e., $\boldvec{\theta}_1 =
\boldvec{\theta}_0$.

\end{enumerate}

\noindent If one takes the proposal density to be symmetric in
$\boldvec{\theta}$ and $\boldvec{\theta}_0$, then this is the {\it
Metropolis}--Hastings algorithm, and the test ratio becomes $\alpha =
\min [1, p(\boldvec{\theta})/p(\boldvec{\theta}_0)]$.  That is, if the
proposed $\boldvec{\theta}$ is at a value of probability higher than
$\boldvec{\theta}_0$, the step is taken.  If the proposed step is
rejected, hop in place.

Methods for assessing and optimizing the performance of the algorithm
are discussed, for example, in Refs.~\cite{Robert04,Liu01,Neal93}. One can, for
example, examine the autocorrelation as a function of the lag $k$,
i.e., the correlation of a sampled point with one $k$ steps removed.
This should decrease as quickly as possible for increasing $k$.
Generally one chooses the proposal density so as to optimize some
quality measure such as the autocorrelation.  For certain problems it
has been shown that one achieves optimal performance when the
acceptance fraction, that is, the fraction of points with $u \le
\alpha$, is around 40\%.  This can be adjusted by varying the width of
the proposal density.  For example, one can use for the proposal pdf a
multivariate Gaussian with the same covariance matrix as that of the
target pdf, but scaled by a constant.

For our example above, MCMC was used to generate points according
to the posterior pdf $p(\theta_0, \theta_1)$ by using a Gaussian
proposal density.  The result is shown in Fig.~\ref{fig:mcmc}.

\setlength{\unitlength}{1.0 cm}
\renewcommand{\baselinestretch}{0.9}
\begin{figure}[htbp]
\begin{picture}(10.0,5)
\put(0.,0){\includegraphics{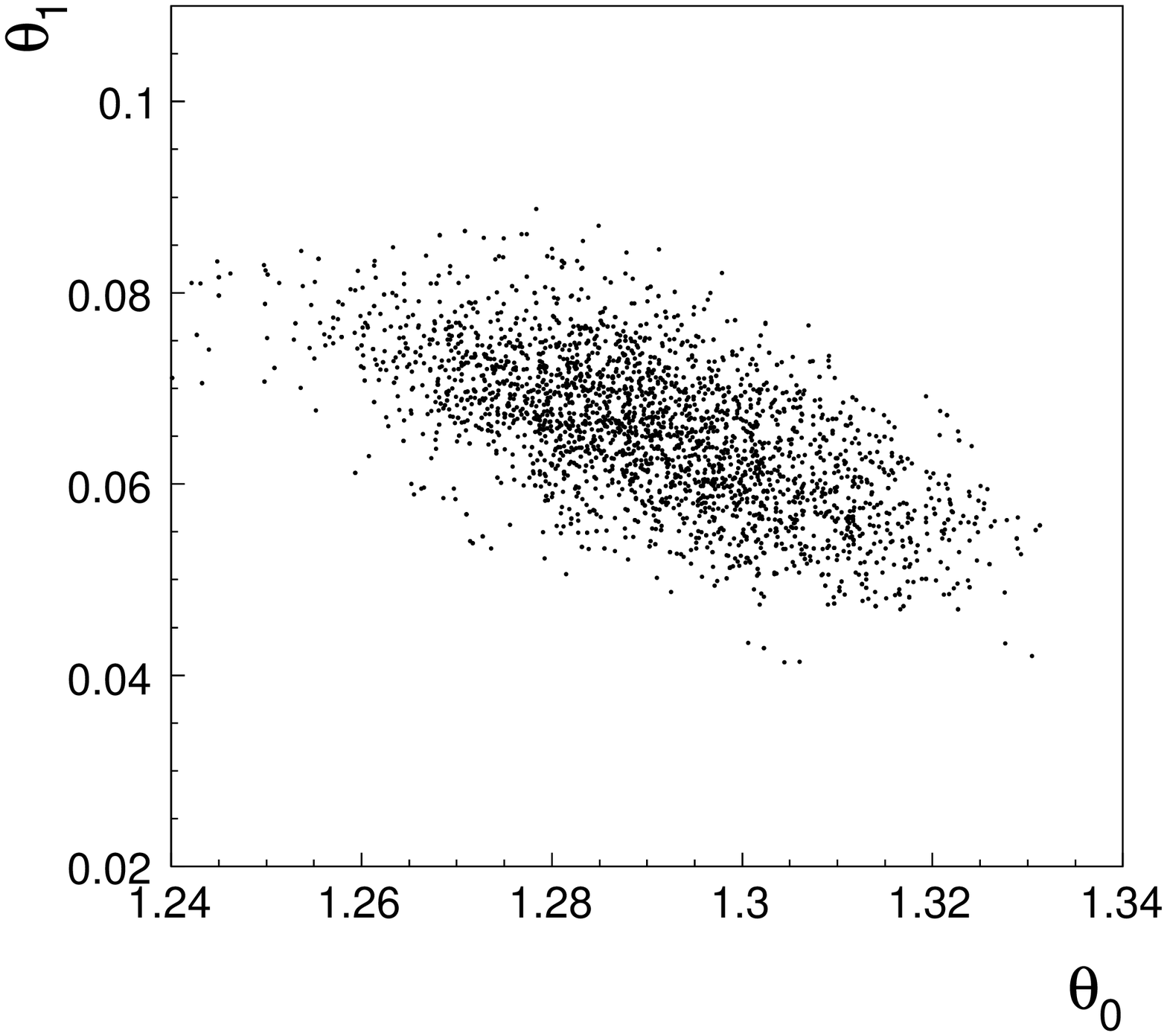}}
\put(5.3,0){\includegraphics{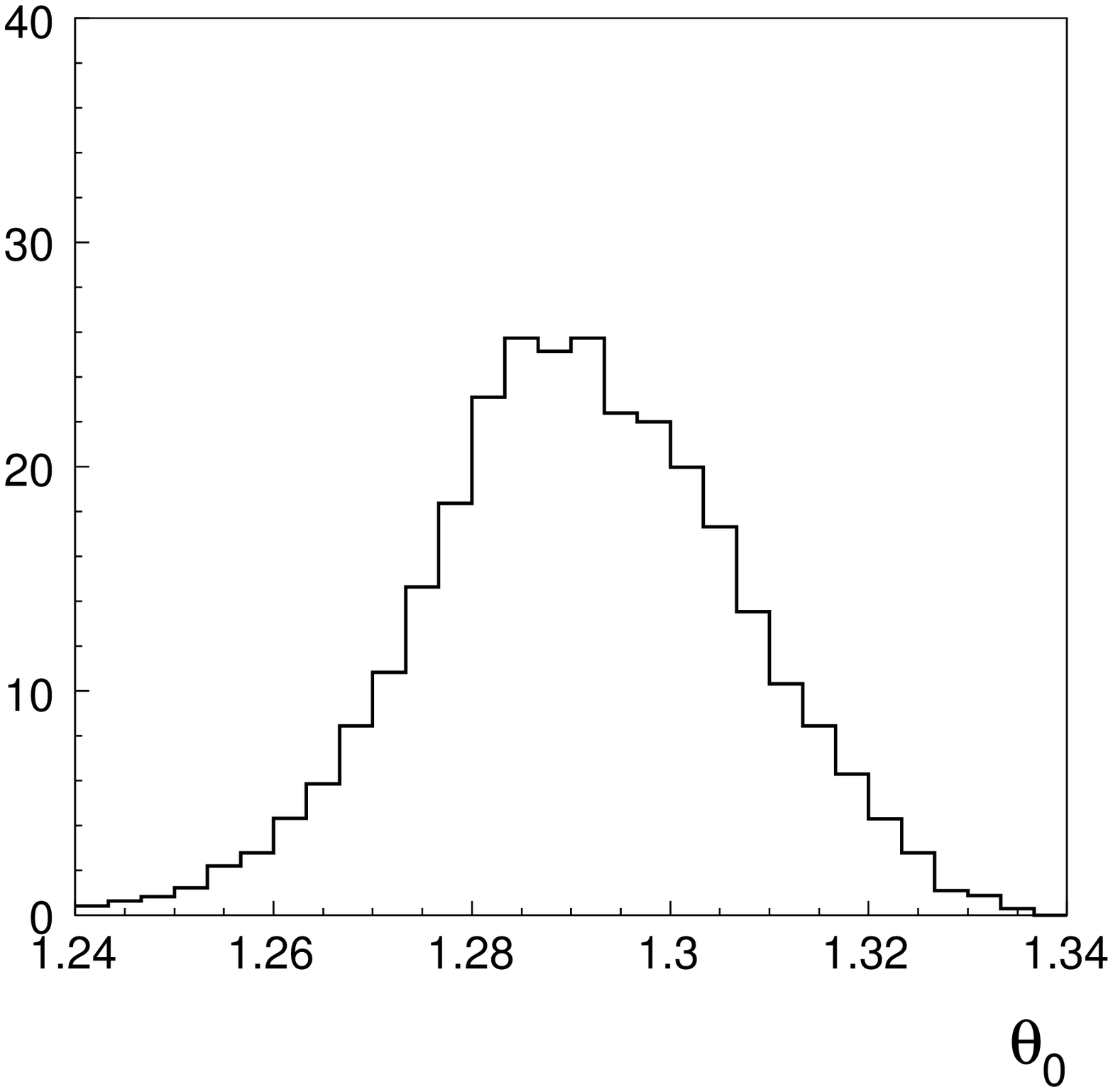}}
\put(10.8,0){\includegraphics{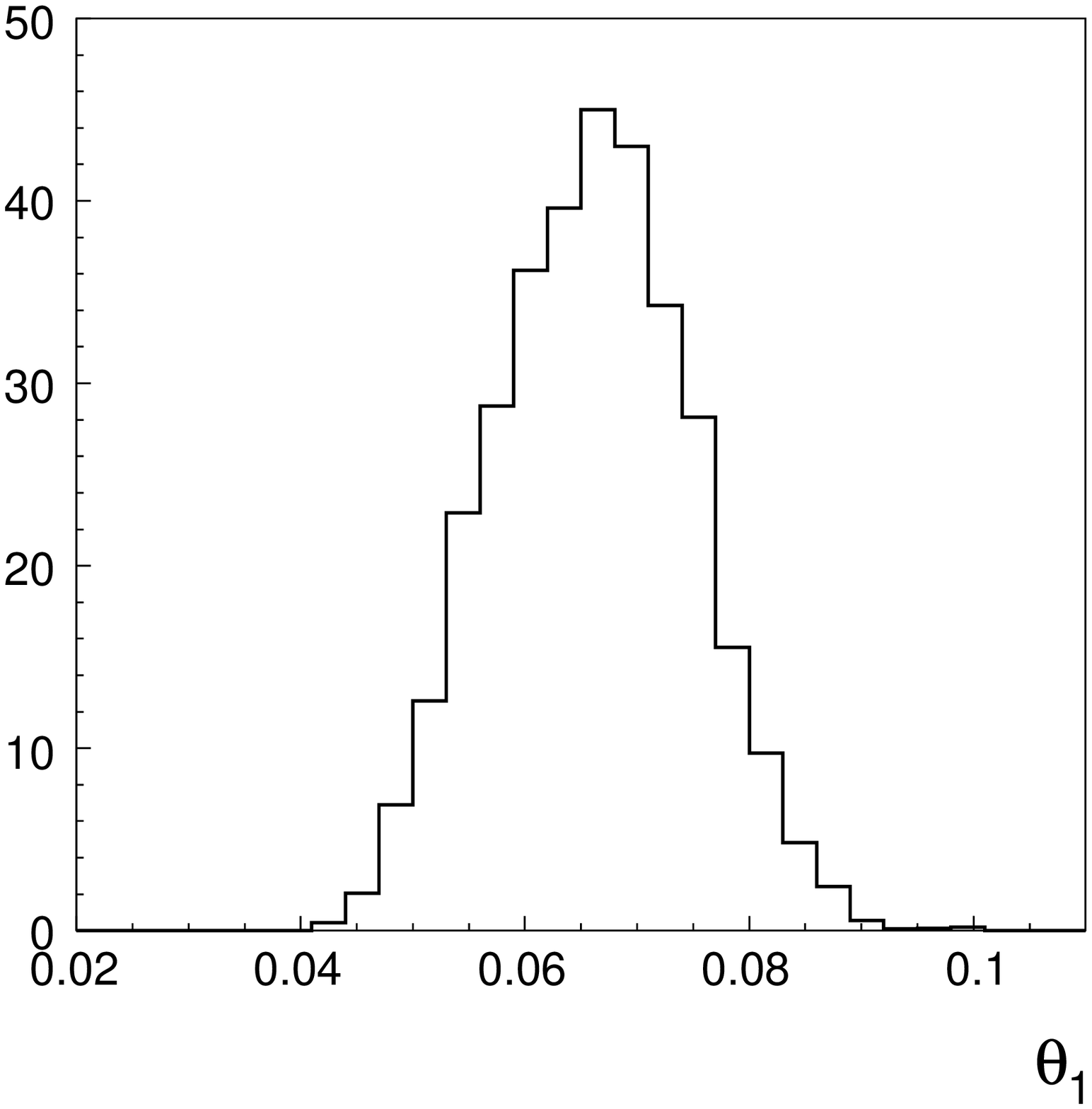}}
\put(4.3,4.){(a)}
\put(9.6,4.){(b)}
\put(14.9,4.){(c)}
\end{picture}
\caption{\small MCMC marginalization of the posterior pdf
$p(\theta_0, \theta_1 | \boldvec{y})$:  (a) scatter-plot of points
in $(\theta_0, \theta_1)$ plane and the marginal distribution of
(b) the parameter of interest $\theta_0$ and (c) the nuisance
parameter $\theta_1$.}
\label{fig:mcmc}
\end{figure}
\renewcommand{\baselinestretch}{1}
\small\normalsize

From the $(\theta_0, \theta_1)$ points in the scatter plot in
Fig.~\ref{fig:mcmc}(a) we simply look at the distribution of the
parameter of interest, $\theta_0$ [Fig.~\ref{fig:mcmc}(b)].  The
standard deviation of this distribution is what we would report as the
statistical error in our measurement of $\theta_0$.  The distribution
of the nuisance parameter $\theta_1$ from Fig.~\ref{fig:mcmc}(c) is
not directly needed, although it may be of interest in some other
context where that parameter is deemed interesting.

In fact one can go beyond simply summarizing the width of the
distributions with the a statistic such as the standard deviation.
The full form of the posterior distribution of $\theta_0$ contains
useful information about where the parameter's true value is likely to
be.  In this example the distributions will in fact turn out to be
Gaussian, but in a more complex analysis there could be non-Gaussian
tails and this information can be relevant in drawing conclusions from
the result.

\subsubsection{Sensitivity analysis}
\label{sec:senanl}

The posterior distribution of $\theta_0$ obtained above encapsulates
all of the analyst's knowledge about the parameter in the light of the
data, given that the prior beliefs were reflected by the density
$\pi(\theta_0, \theta_1)$.  A different analyst with different prior
beliefs would in general obtain a different posterior pdf.  We would
like the result of a Bayesian analysis to be of value to the broader
scientific community, not only to those that share the prior beliefs
of the analyst.  And therefore it is important in a Bayesian analysis
to show by how much the posterior probabilities would change upon some
reasonable variation in the prior.  This is sometimes called the {\it
sensitivity analysis} and is an important part of any Bayesian
calculation.

In the example above, we can imagine a situation where there was no
prior measurement $t_1$ of the parameter $\theta_1$, but rather a
theorist had told us that, based on considerations of symmetry,
consistency, aesthetics, etc., $\theta_1$ was ``almost
certainly'' positive, and had a magnitude ``probably less than 0.1 or
so''.  When pressed to be precise, the theorist sketches a curve
roughly resembling an exponential with a mean of 0.1.  So we can
express this prior as

\begin{equation}
\label{eq:pi1exp}
\pi_1(\theta_1) = \frac{1}{\tau} e^{-\theta_1/\tau} \quad (\theta_1 \ge 0) 
\;,
\end{equation}

\noindent with $\tau \approx 0.1$.  We can substitute this prior into
Bayes' theorem~(\ref{eq:bthmlf}) to obtain the joint pdf for
$\theta_0$ and $\theta_1$, and then marginalize to find the pdf for
$\theta_0$.  Doing this numerically with MCMC results in the posterior
distributions shown in Fig.~\ref{fig:PiExp}(a).

\setlength{\unitlength}{1.0 cm}
\renewcommand{\baselinestretch}{0.9}
\begin{figure}[htbp]
\begin{picture}(10.0,6.5)
\put(0.5,0){\includegraphics{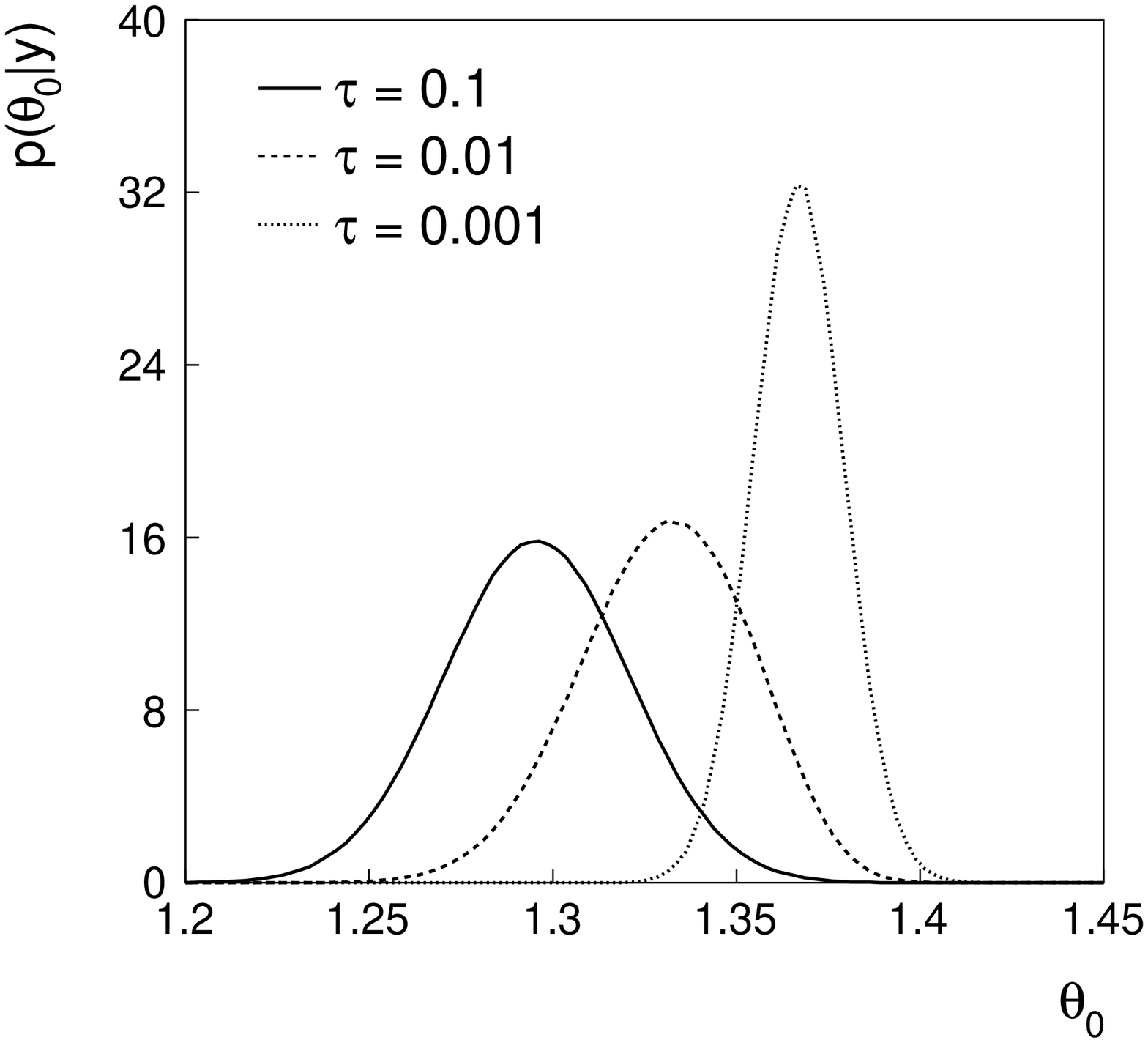}}
\put(8.5,0){\includegraphics{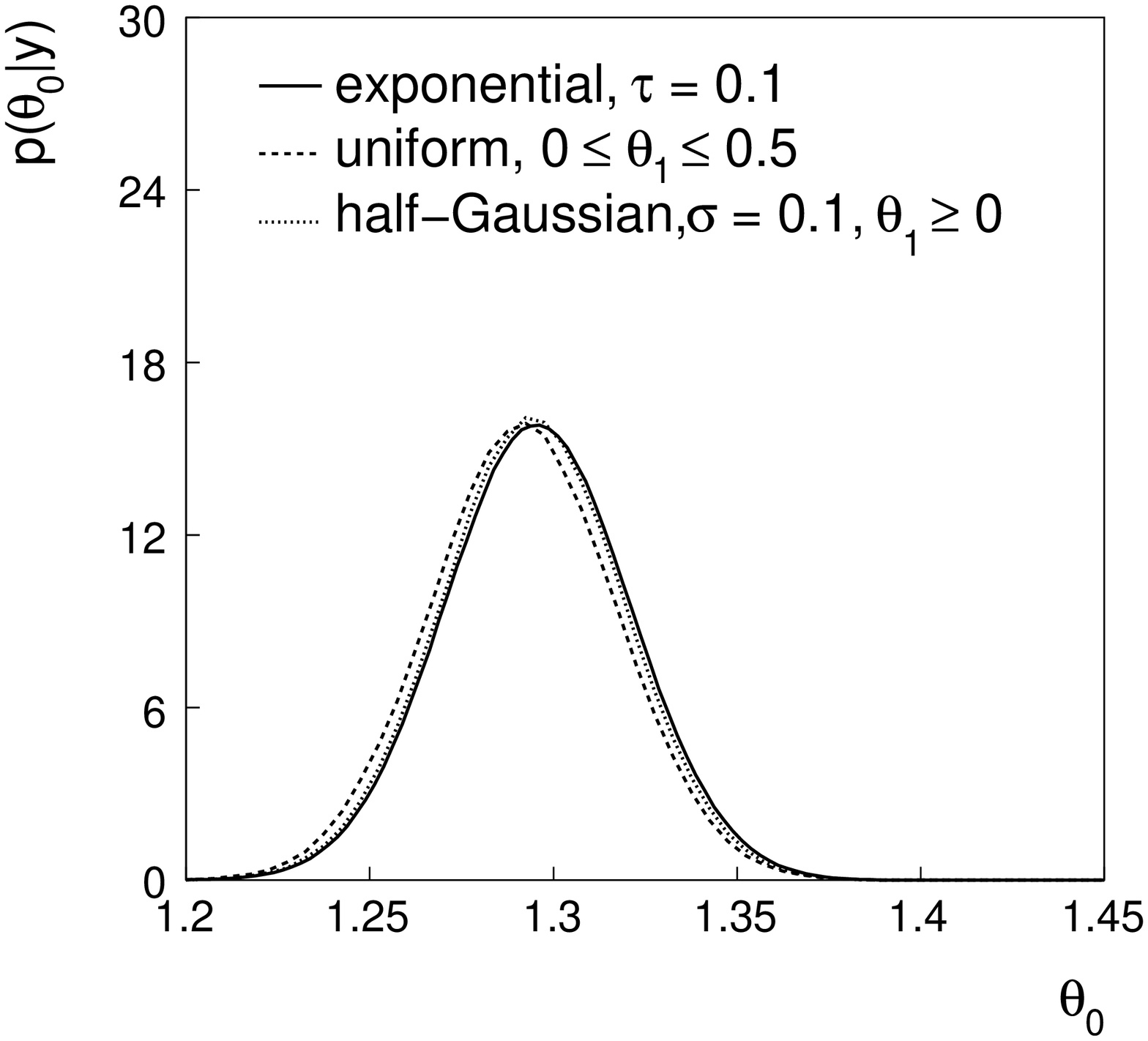}}
\put(0.,5.5){(a)}
\put(15.5,5.5){(b)}
\end{picture}
\caption{\small Posterior probability densities for the parameter
$\theta_0$ obtained using (a) an exponential prior for $\theta_0$
of different widths and (b) several different functional forms
for the prior.}
\label{fig:PiExp}
\end{figure}
\renewcommand{\baselinestretch}{1}
\small\normalsize

Now the theorist who proposed this prior for $\theta_1$ may feel
reluctant to be pinned down, and so it is important to recall (and
to reassure the theorist about) the ``if-then'' nature of a Bayesian
analysis.  One does not have to be absolutely certain about the prior
in Eq.~(\ref{eq:pi1exp}).  Rather, Bayes' theorem simply says that
{\it if} one were to have these prior beliefs, {\it then} we obtain
certain posterior beliefs in the light of the data.

One simple way to vary the prior here is to try different values of
the mean $\tau$, as shown in Fig.~\ref{fig:PiExp}(a).  We see here the
same basic feature as shown already in the frequentist analysis,
namely, that when one increases the precision about the nuisance
parameter, $\theta_1$, then the knowledge about the parameter of
interest, $\theta_0$, is improved.

Alternatively (or in addition) we may try different functional forms
for the prior, as shown in Fig.~\ref{fig:PiExp}(b).  In this case
using a uniform distribution for $\pi_1(\theta_1)$ with $0 \le
\theta_1 \le 0.5$ or Gaussian with $\sigma = 0.1$ truncated for
$\theta_1 < 0$ both give results similar to the exponential with a
mean of $0.1$.  So one concludes that the result is relatively
insensitive to the detailed nature of the tails of $\pi_1(\theta_1)$.

\subsection{A fit with systematic errors}
\label{sec:fitsys}

We can now generalize the example of Section~\ref{sec:linefit} to explore
some further aspects of a Bayesian analysis.  Let us suppose that we
are given a set of $n$ measurements as above, but now in addition to
the statistical errors we also are given systematic errors.  That is,
we are given $y_i \pm \sigma_i^{\rm stat} \pm \sigma_i^{\rm sys}$ for
$i = 1, \ldots, n$ where the measurements as before are each carried
out for a specified value of a control variable $x$.

More generally, instead of having $y_i \pm \sigma_i^{\rm stat} \pm
\sigma_i^{\rm sys}$ it may be that the set of measurements comes with
an $n \times n$ covariance matrix $V^{\rm stat}$ corresponding to the
statistical errors and another matrix $V^{\rm sys}$ for the systematic
ones.  Here the square roots of the diagonal elements give the errors
for each measurement, and the off-diagonal elements provide
information on how they are correlated.

As before we assume some functional form $\mu(x; \boldvec{\theta})$
for the expectation values of the $y_i$.  This could be the linear
model of Eq.~(\ref{eq:muofx}) or something more general, but in any
case it depends on a vector of unknown parameters $\boldvec{\theta}$.
In this example, however, we will allow that the model is not perfect,
but rather could have a systematic bias.  That is, we write that the true
expectation value of the $i$th measurement can be written

\begin{equation}
\label{eq:expofy}
E[y_i] = \mu(x_i; \boldvec{\theta}) + b_i \;,
\end{equation}

\noindent where $b_i$ represents the bias.  The $b_i$ can be viewed as
the systematic errors of the model, present even when the parameters
$\boldvec{\theta}$ are adjusted to give the best description of the
data.  We do not know the values of the $b_i$.  If we did, we would
account for them in the model and they would no longer be biases.  We
do not in fact know that their values are nonzero, but we are allowing
for the possibility that they could be.  The reported systematic
errors are intended as a quantitative measure of how large we
expect the biases to be.

As before, the goal is to make inferences about the parameters
$\boldvec{\theta}$; some of these may be of direct interest and others
may be nuisance parameters.  In Section~\ref{sec:genfreq} we will try to
do this using the frequentist approach, and in Section~\ref{sec:genbay}
we will use the Bayesian method.

\subsubsection{A frequentist fit with systematic errors}
\label{sec:genfreq}

If we adopt the frequentist approach, we need to write down a
likelihood function such as Eq.~(\ref{eq:yilikelihood}), but here we
know in advance that the model $\mu(x; \boldvec{\theta})$ is not expected
to be fully accurate.  Furthermore it is not clear how to insert the
systematic errors.  Often, perhaps without a clear justification, one
simply adds the statistical and systematic errors in quadrature, or in
the case where one has the covariance matrices $V^{\rm stat}$ and
$V^{\rm sys}$, they are summed to give a sort of `full' covariance
matrix:

\begin{equation}
\label{eq:fullcov}
V_{ij} = V_{ij}^{\rm stat} + V_{ij}^{\rm sys} \;.
\end{equation}

\noindent One might then use this in a multivariate Gaussian
likelihood function, or equivalently it could be used to construct the
$\chi^2$,

\begin{equation}
\label{eq:chi2multi}
\chi^2(\boldvec{\theta}) = (\boldvec{y} - \boldvec{\mu}(\boldvec{\theta}))^T
V^{-1} (\boldvec{y} - \boldvec{\mu}(\boldvec{\theta})) \;,
\end{equation}

\noindent which is then minimized to find the LS estimators for
$\boldvec{\theta}$.  In Eq.~(\ref{eq:chi2multi}) the vector
$\boldvec{y} = (y_1, \ldots, y_n)$ should be understood as a column
vector, $\boldvec{\mu}(\boldvec{\theta}) = (\mu(x_1;
\boldvec{\theta}), \ldots, \mu(x_n; \boldvec{\theta}))$ is the
corresponding vector of model values, and the superscript $T$
represents the transpose (row) vector.  Minimizing this $\chi^2$ gives
the generalized LS estimators $\hat{\boldvec{\theta}}$, and the usual
procedures can be applied to find their covariances, which now in some
sense include the systematics.

But in what sense is there any formal justification for adding the
covariance matrices in Eq.~(\ref{eq:fullcov})?  Next we will treat
this problem in the Bayesian framework and see that there is indeed
some reason behind this recipe, but with limitations, and further we
will see how to get around these limitations.

\subsubsection{The equivalent Bayesian fit}
\label{sec:genbay}

In the corresponding Bayesian analysis, one treats the statistical
errors as given by $V^{\rm stat}$ as reflecting the distribution of
the data $\boldvec{y}$ in the likelihood.  The systematic errors,
through $V^{\rm sys}$, reflect the width of the prior probabilities
for the bias parameters $b_i$.  That is, we take

\begin{eqnarray}
\label{eq:equivbayes}
L(\boldvec{y} | \boldvec{\theta}, \boldvec{b}) & \propto & \exp \left[ - 
\begin{matrix} \frac{1}{2} \end{matrix}
(\boldvec{y} - \boldvec{\mu}(\boldvec{\theta}) - 
\boldvec{b})^T V_{\rm stat}^{-1}
(\boldvec{y} - \boldvec{\mu}(\boldvec{\theta}) - 
\boldvec{b}) \right] \;, \\*[0.2 cm]
\label{eq:pib}
\pi_b(\boldvec{b}) & \propto & 
\exp \left[ - \begin{matrix} \frac{1}{2} \end{matrix}
\boldvec{b}^T V_{\rm sys}^{-1} \boldvec{b} \right] \;,
 \quad  \quad
\pi_{\theta}(\boldvec{\theta}) = \mbox{const.} \;, \\*[0.2 cm] 
\label{eq:bayesthetab}
p(\boldvec{\theta}, \boldvec{b} | \boldvec{y}) & 
\propto & L(\boldvec{y} | \boldvec{\theta},
\boldvec{b}) \pi_{\theta}(\boldvec{\theta}) \pi_{b}(\boldvec{b}) \;,
\end{eqnarray}

\noindent where in~(\ref{eq:bayesthetab}), Bayes' theorem is used to
obtain the joint probability for the parameters of interest,
$\boldvec{\theta}$, and also the biases $\boldvec{b}$.  To obtain the
probability for $\boldvec{\theta}$ we integrate (marginalize) over
$\boldvec{b}$,

\begin{equation}
\label{eq:margprob}
p(\boldvec{\theta} | \boldvec{y} ) = 
\int p(\boldvec{\theta}, \boldvec{b} | \boldvec{y}) 
\, d \boldvec{b} \;.
\end{equation}

\noindent One finds that the mode of $p(\boldvec{\theta} |
\boldvec{y})$ is at the same position as the least-squares estimates,
and its covariance will be the same as obtained from the frequentist
analysis where the full covariance matrix was given by the sum $V =
V^{\rm stat} + V^{\rm sys}$.  So this can be taken in effect as the
formal justification for the addition in quadrature of statistical and
systematic errors in a least-squares fit.

\subsubsection{The error on the error}
\label{sec:eoe}

If one stays with the prior probabilities used above, the Bayesian and
least-squares approaches deliver essentially the same result.  An
advantage of the Bayesian framework, however, is that it allows one to
refine the assessment of the systematic uncertainties as expressed
through the prior probabilities.

For example, the least-squares fit including systematic errors is
equivalent to the assumption of a Gaussian prior for the biases.  A
more realistic prior would take into account the experimenter's own
uncertainty in assigning the systematic error, i.e., the `error on the
error'.  Suppose, for example, that the $i$th measurement is
characterized by a reported systematic uncertainty $\sigma^{\rm
sys}_i$ and an unreported factor $s_i$, such that the prior for the
bias $b_i$ is

\begin{equation}
\label{eq:pieoe}
\pi_{b}(b_i) = \int \frac{1}{\sqrt{2 \pi} s_i \sigma^{\rm sys}_i }
\exp \left[ - \frac{1}{2}
\frac{b_i^2}{(s_i \sigma^{\rm sys}_i)^2} \right]
\pi_s(s_i) \, ds_i \;.
\end{equation}

\noindent Here the `error on the error' is encapsulated in the prior
for the factor $s$, $\pi_s(s)$. For this we can take whatever function
is deemed appropriate.  For some types of systematic error it could be
close to the ideal case of a delta function centred about unity.  Many
reported systematics are, however, at best rough guesses, and one
could easily imagine a function $\pi_{s}(s)$ with a mean of unity but
a standard deviation of, say, $0.5$ or more.  Here we show examples
using a Gamma distribution for $\pi_s(s)$, which results in
substantially longer tails for the prior $\pi_b(b)$ than those of the
Gaussian.  This can be seen in Fig.~\ref{fig:Pib}, which shows $\ln
\pi_b(b)$ for different values of the standard deviation of
$\pi_{s}(s)$, $\sigma_s$.  Related studies using an inverse Gamma
distribution can be found in Refs.~\cite{dagostini,dose}, which have the
advantage that the posterior pdf can be written down in closed form.

\setlength{\unitlength}{1.0 cm}
\renewcommand{\baselinestretch}{0.8}
\begin{figure}[htbp]
\begin{picture}(10.0,6)
\put(1.,-0.2){\includegraphics{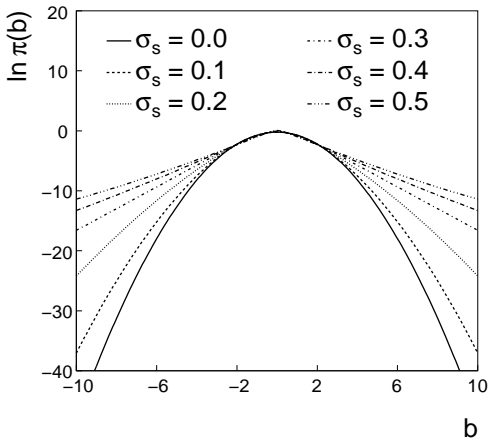}}
\put(9.0,.8){\makebox(6,4)[b]{\begin{minipage}[b]{6cm}
\protect\caption{{\small The log of the prior pdf for a 
bias parameter $b$ for different values of the standard deviation 
of $\pi_s(s)$.}
\protect\label{fig:Pib}}
\end{minipage}}}
\end{picture}
\end{figure}
\renewcommand{\baselinestretch}{1}
\small\normalsize

Using a prior for the biases with tails longer than those of a
Gaussian results in a reduced sensitivity to outliers, which arise
when an experimenter overlooks an important source of systematic
uncertainty in the estimated error of a measurement.  As a simple test
of this, consider the sample data shown in Fig.~\ref{fig:outlier}(a).
Suppose these represent four independent measurements of the same
quantity, here a parameter called $\mu$, and the goal is to combine
the measurements to provide a single estimate of $\mu$.  That is, we
are effectively fitting a horizontal line to the set of measured $y$
values, where the control variable $x$ is just a label for the
measurements.

In this example, suppose that each measurement $y_i$, $i=1, \ldots 4$,
is modelled as Gaussian distributed about $\mu$, having a standard
deviation $\sigma_{\rm stat} = 0.1$, and furthermore each measurement
has a systematic uncertainty $\sigma_{\rm sys} = 0.1$, which here is
taken to refer to the standard deviation of the Gaussian component of
the prior $\pi_{b}(b_i)$.  This is then folded together with
$\pi_s(s_i)$ to get the full prior for $b_i$ using
Eq.~(\ref{eq:pieoe}), and the joint prior for the vector of bias
parameters is simply the product of the corresponding terms, as the
systematic errors here are treated as being independent.  These
ingredients are then assembled according to the recipe of
Eqs.~(\ref{eq:equivbayes})--(\ref{eq:margprob}) to produce the
posterior pdf for $\mu$, $p(\mu | \boldvec{y})$.

Results of the exercise are shown in Fig.~\ref{fig:outlier}.  In
Fig.~\ref{fig:outlier}(a), the four measurements $y_i$ are reasonably
consistent with each other.  Figure~\ref{fig:outlier}(b) shows the
corresponding posterior $p(\mu | \boldvec{y})$ for two values of
$\sigma_{s}$, which reflect differing degrees of non-Gaussian tails in
the prior for the bias parameters, $\pi_{b}(b_i)$.  For $\sigma_{s} =
0$, the prior for the bias is exactly Gaussian, whereas for
$\sigma_{s} = 0.5$, the non-Gaussian tails are considerably longer, as
can be seen from the corresponding curves in Fig.~\ref{fig:Pib}.
The posterior pdfs for both cases are almost identical, as can be see
in Fig.~\ref{fig:outlier}(a).  Determining the mean and standard
deviation of the posterior for each gives $\hat{\mu} = 1.000 \pm 0.71$
for the case of $\sigma_{s} = 0$, and $\hat{\mu} = 1.000 \pm 0.72$
for $\sigma_{s} = 0.5$.  So assuming a 50\% ``error on the error''
here one only inflates the error of the averaged result by a small
amount.

\setlength{\unitlength}{1.0 cm}
\renewcommand{\baselinestretch}{0.9}
\begin{figure}[htbp]
\begin{picture}(10.0,8.5)
\put(1.5,5.5){\includegraphics{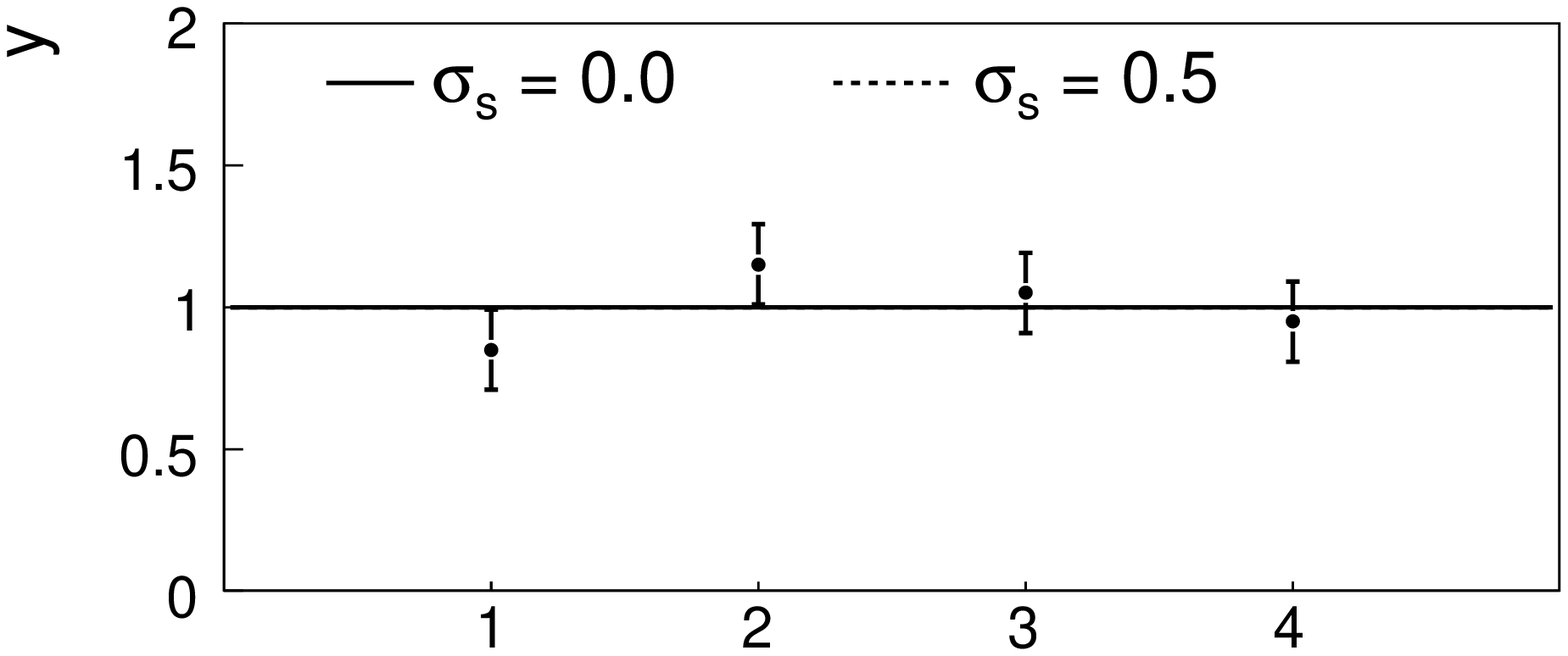}}
\put(8.5,5.5){\includegraphics{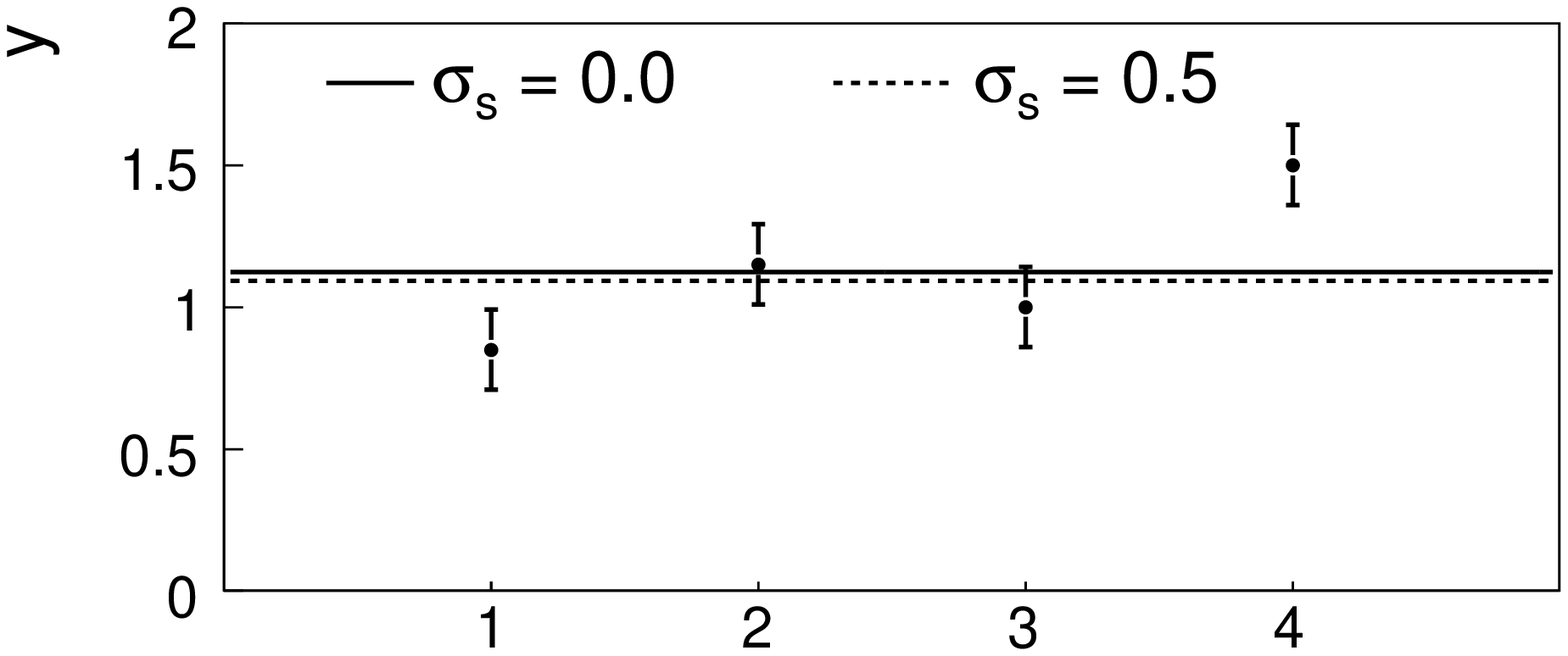}}
\put(0.5,7.5){(a)}
\put(15.,7.5){(b)}
\put(1.5,0){\includegraphics{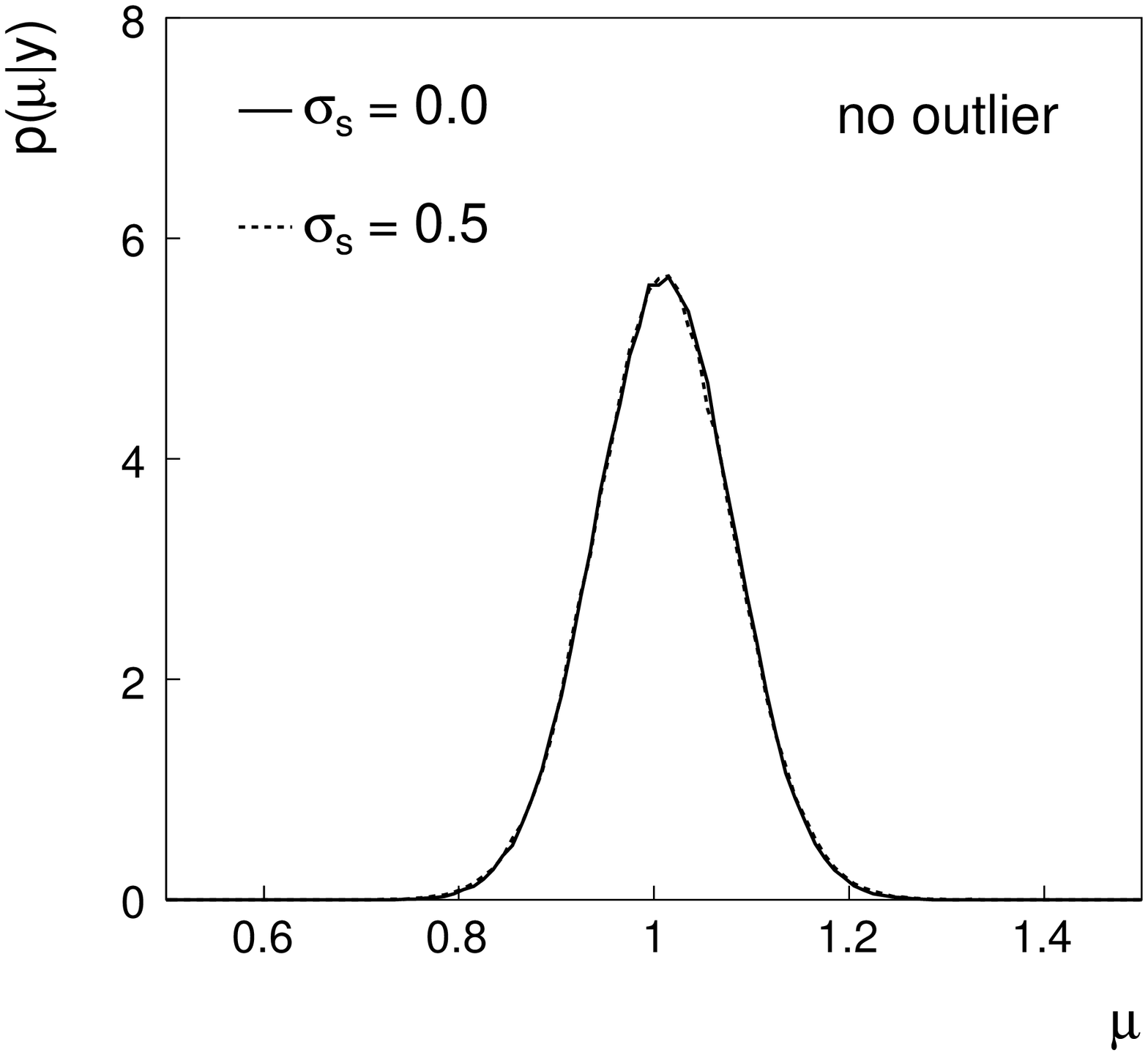}}
\put(8.5,0){\includegraphics{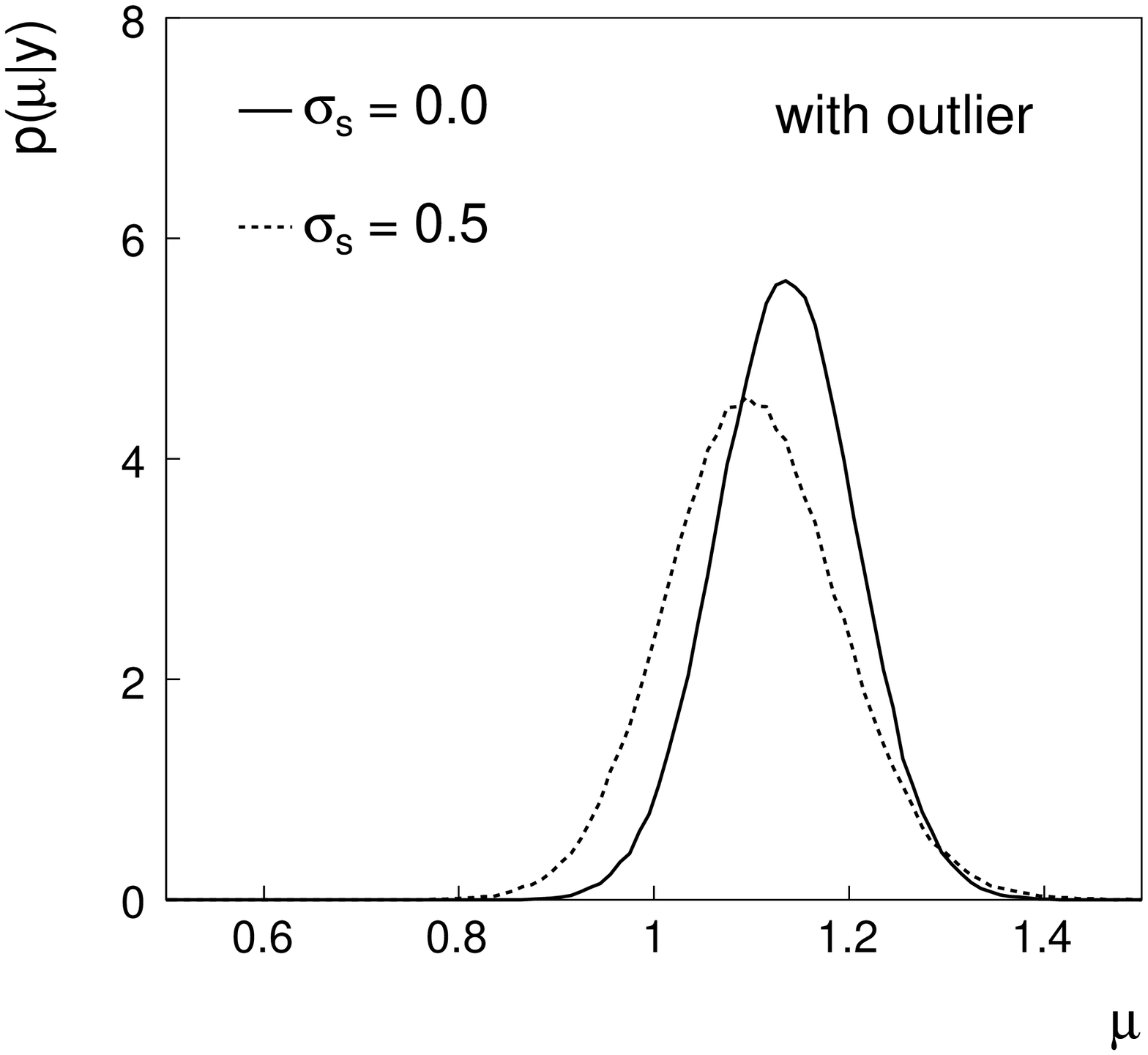}}
\put(0.5,3.5){(c)}
\put(15.,3.5){(d)}
\end{picture}
\caption{\small (a) Data values which are relatively consistent and
(b) a data set with an outlier; the horizontal lines indicate the
posterior mean for two different values of the parameter $\sigma_{s}$.
(c) and (d) show the posterior distributions corresponding to (a) and
(b), respectively. (The dashed and solid curves in (a) and (c)
overlap.)}
\label{fig:outlier}
\end{figure}
\renewcommand{\baselinestretch}{1}
\small\normalsize

Now consider the case where one of the measured values is
substantially different from the other three, as shown in
Fig.~\ref{fig:outlier}(c).  Here using the same priors for the bias
parameters results in the posteriors shown in
Fig.~\ref{fig:outlier}(d).  The posterior means and standard
deviations are $\hat{\mu} = 1.125 \pm 0.71$ for the case of
$\sigma_{s} = 0$, and $\hat{\mu} = 1.093 \pm 0.089$ for $\sigma_{s}
= 0.5$.

When we assume a purely Gaussian prior for the bias ($\sigma_s =
0.0$), the presence of the outlier has in fact no effect on the width
of the posterior.  This is rather counter-intuitive and results from
our assumption of a Gaussian likelihood for the data and a Gaussian
prior for the bias parameters.  The posterior mean is however pulled
substantially higher than the three other measurements, which are
clustered around $1.0$.  If the priors $\pi_{b}(b_i)$ have longer
tails, as occurs when we take $\sigma_{s} = 0.5$, then the posterior
is broader, and furthermore it is pulled less far by the outlier, as
can be seen in Fig.~\ref{fig:outlier}(d).

The fact is that the width of the posterior distribution, which
effectively tells us the uncertainty on the parameter of interest
$\mu$, becomes coupled to the internal consistency of the data.  In
contrast, in the (frequentist) least-squares method, or in the
Bayesian approach using a Gaussian prior for the bias parameters, the
final uncertainty on the parameter of interest is unaffected by the
presence of outliers.  And in many cases of practical interest, it
would be in fact appropriate to conclude that the presence of outliers
should indeed increase one's uncertainty about the final parameter
estimates.  The example shown here can be generalized to cover a wide
variety of model uncertainties by including prior probabilities for an
enlarged set of model parameters.


\subsection{Summary on Bayesian methods}
\label{sec:sumbayes}

In these lectures we have seen how Bayesian methods can be used in
parameter estimation, and this has also given us the opportunity to
discuss some aspects of Bayesian computation, including the important
tool of Markov Chain Monte Carlo.  Although Bayesian and frequentist
methods may often deliver results that agree numerically, there is an
important difference in their interpretation.  Furthermore, Bayesian
methods allow one to incorporate prior information that may be based
not on other measurements but rather on theoretical arguments or
purely subjective considerations.  And as these considerations may not
find universal agreement, it is important to investigate how the
results of a Bayesian analysis would change for a reasonable variation
of the prior probabilities.

It is important to keep in mind that in the Bayesian approach, all
information about the parameters is encapsulated in the posterior
probabilities.  So if the analyst also wants to set upper limits or
determine intervals that cover the parameter with a specified
probability, then this is a straightforward matter of finding the
parameter limits such that the integrated posterior pdf has the
desired probability content.  A discussion of Bayesian methods to the
important problem of setting upper limits on a Poisson parameter is
covered in Ref.~\cite{PDG} and references therein; we will not have time
in these lectures to go into that question here.

We will also unfortunately not have time to explore Bayesian model
selection.  This allows one to quantify the degree to which the the
data prefer one model over the other using a quantity called the Bayes
factor.  These have not yet been widely used in particle physics but
should be kept in mind as providing important complementary
information to the corresponding outputs of frequentist hypothesis
testing such as $p$-values.  A brief description of Bayes factors can
be found in Ref.~\cite{PDG} and a more in-depth treatment is given in
Ref.~\cite{KassRaftery95}.

\section{Topics in multivariate analysis}
\label{sec:multivariate}

In the second part of these lectures we will take a look at the
important topic of multivariate analysis.  In-depth information on
this topic can be found in the textbooks~\cite{Bishop,Hastie,Duda,Webb}.  In a particle physics context,
multivariate methods are often used when selecting events of a certain
type using some potentially large number of measurable characteristics
for each event.  The basic framework we will use to examine these
methods is that of a frequentist hypothesis test.

The fundamental unit of data in a particle physics experiment is the
`event', which in most cases corresponds to a single particle
collision.  In some cases it could be instead a decay, and the picture
does not change much if we look, say, at individual particles or
tracks.  But to be concrete let us suppose that we want to search for
events from proton--proton collisions at the LHC that correspond to
some interesting `signal' process, such as supersymmetry.

When running at full intensity, the LHC should produce close to a
billion events per second.  After a quick sifting, the data from
around 200 per second are recorded for further study, resulting in
more than a billion events per year.  But only a tiny fraction of
these are of potential interest.  If one of the speculative theories
such as supersymmetry turns out to be realized in Nature, then this
will result in a subset of events having characteristic features, and
the SUSY events will simply be mixed in randomly with a much larger
number of Standard Model events.  The relevant distinguishing features
depend on what new physics Nature chooses to reveal, but one might
see, for example, high $p_{\rm T}$ jets, leptons, missing energy.

Unfortunately, background processes (e.g., Standard Model events) can
often mimic these features and one will not be able to say with
certainty that a given event shows a clear evidence for something new
such as supersymmetry.  For example, even Standard Model events can
contain neutrinos which also escape undetected.  The typical amount
and pattern of missing energy in these events differs on average,
however, from what a SUSY event would give, and so a statistical
analysis can be applied to test whether something besides Standard
Model events is present.

In a typical analysis there is a class of event we are interested in
finding (signal), and these, if they exist at all, are mixed in with
the rest of the events (background).  The data for each event is some
collection of numbers $\boldvec{x} = (x_1, \ldots, x_n)$ representing
particle energies, momenta, etc.  We will refer to these as the {\it
input variables} of the problem.  And the probabilities are joint
densities for $\boldvec{x}$ given the signal (s) or background (b)
hypotheses: $f(\boldvec{x} | \mbox{s})$ and $f(\boldvec{x} |
\mbox{b})$.

To illustrate the general problem, consider the scatterplots shown in
Fig.~\ref{fig:scatter}.  These show the distribution of two variables,
$x_1$ and $x_2$, which represent two out of a potentially large number
of quantities measured for each event.  The blue circles could
represent the sought after signal events, and the red triangles the
background.  In each of the three figures there is a decision boundary
representing a possible way of classifying the events.

\setlength{\unitlength}{1.0 cm}
\renewcommand{\baselinestretch}{0.9}
\begin{figure}[htbp]
\begin{picture}(10.0,5)
\put(0.,0){\includegraphics{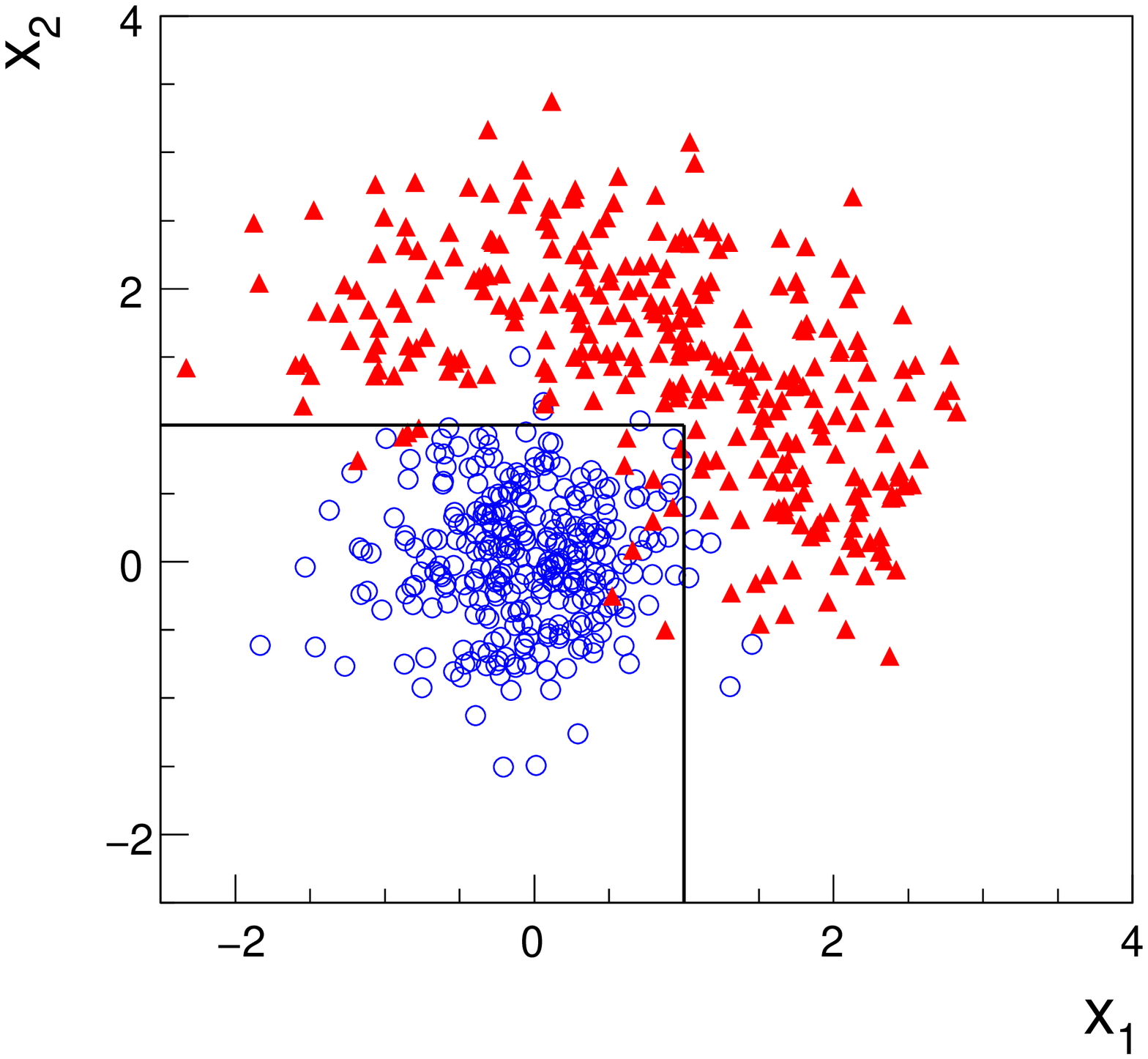}}
\put(5.3,0){\includegraphics{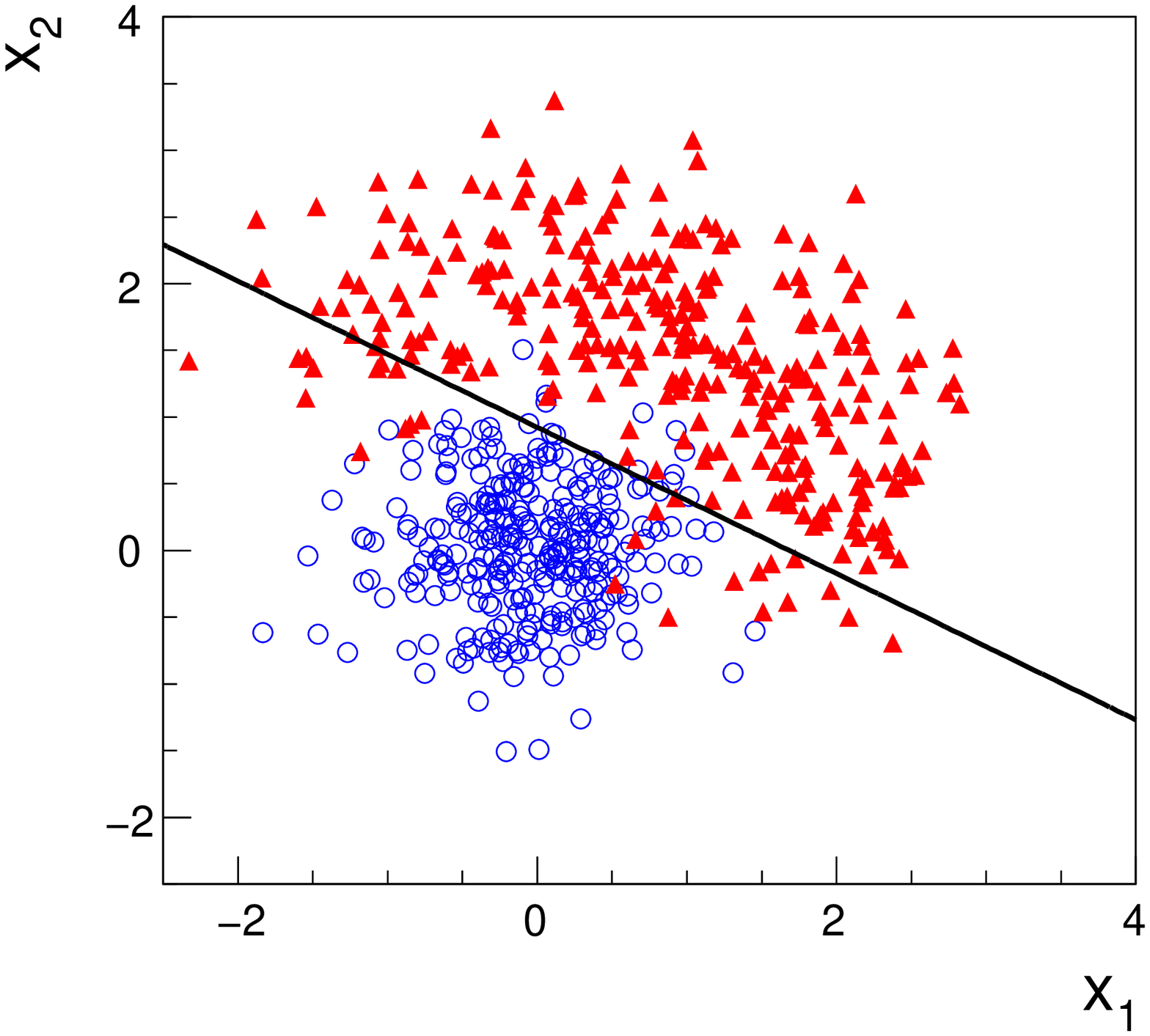}}
\put(10.8,0){\includegraphics{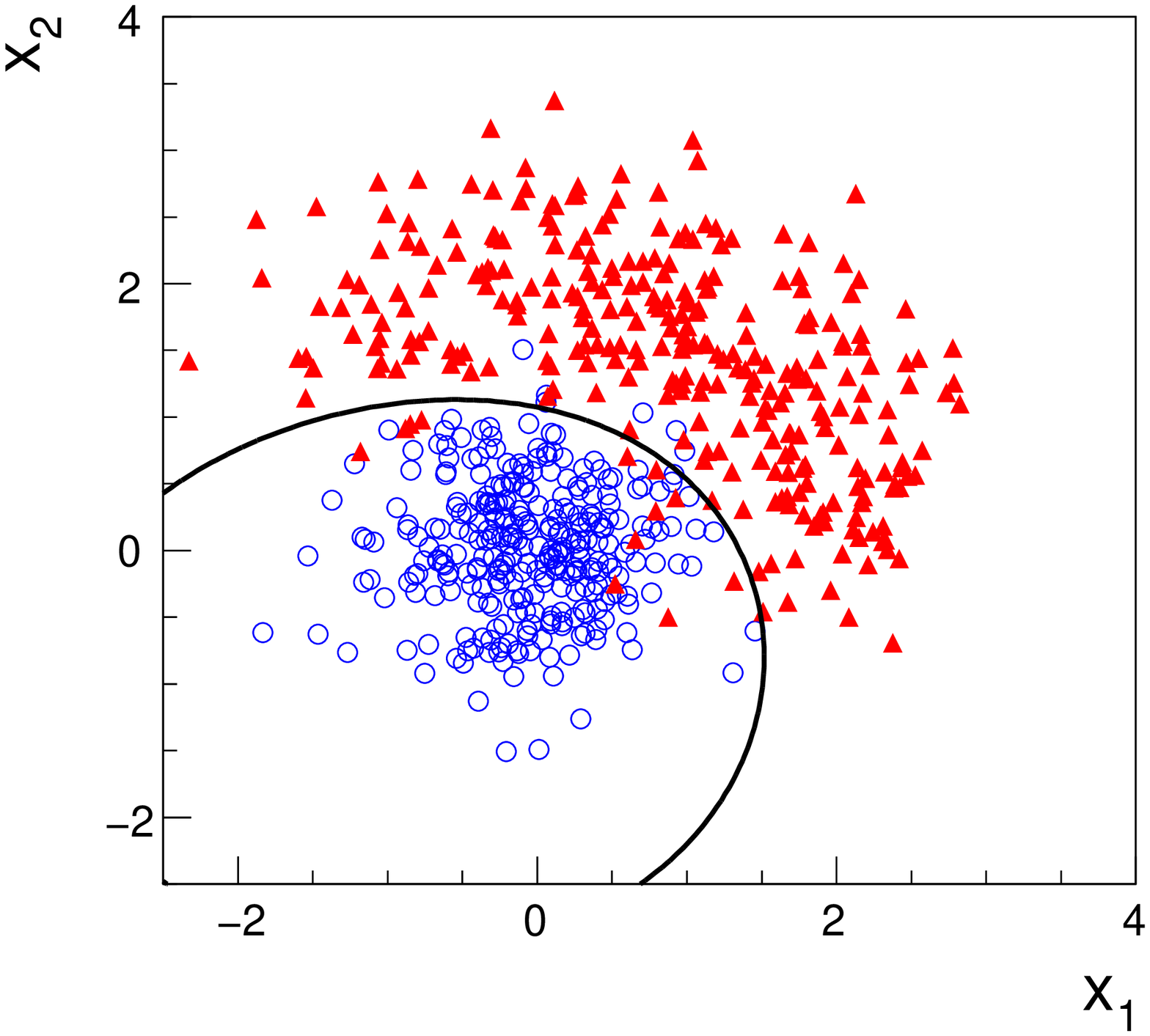}}
\put(4.4,4.1){(a)}
\put(9.6,4.1){(b)}
\put(15.1,4.1){(c)}
\end{picture}
\caption{\small Scatter plots of two variables corresponding
to two hypotheses: signal and background.  Event selection could be
based, e.g., on (a) cuts, (b) a linear boundary, (c) a nonlinear boundary.}
\label{fig:scatter}
\end{figure}
\renewcommand{\baselinestretch}{1}
\small\normalsize

Figure~\ref{fig:scatter}(a) represents what is commonly called the
`cut-based' approach.  One selects signal events by requiring $x_1 <
c_1$ and $x_2 < c_2$ for some suitably chosen cut values $c_1$ and
$c_2$.  If $x_1$ and $x_2$ represent quantities for which one has some
intuitive understanding, then this can help guide one's choice of the
cut values.

Another possible decision boundary is made with a diagonal cut as
shown in Fig.~\ref{fig:scatter}(b).  One can show that for certain
problems a linear boundary has optimal properties, but in
the example here, because of the curved nature of the distributions,
neither the cut-based nor the linear solution is as good as the
nonlinear boundary shown in Fig.~\ref{fig:scatter}(c).

The decision boundary is a surface in the $n$-dimensional space of
input variables, which can be represented by an equation of the form
$y(\boldvec{x}) = y_{\rm cut}$, where $y_{\rm cut}$ is some constant.
We accept events as corresponding to the signal hypothesis if they are
on one side of the boundary, e.g., $y(\boldvec{x}) \le y_{\rm cut}$
could represent the acceptance region and $y(\boldvec{x}) > y_{\rm
cut}$ could be the rejection region.

Equivalently we can use the function $y(\boldvec{x})$ as a scalar {\it
test statistic}.  Once its functional form is specified, we can
determine the pdfs of $y(\boldvec{x})$ under both the signal and
background hypotheses, $p(y|\mbox{s})$ and $p(y|\mbox{b})$.  The
decision boundary is now effectively a single cut on the scalar
variable $y$, as illustrated in Fig.~\ref{fig:TestStat}.

\setlength{\unitlength}{1.0 cm}
\renewcommand{\baselinestretch}{0.8}
\begin{figure}[htbp]
\begin{picture}(10.0,5)
\put(1.5,0.){\includegraphics{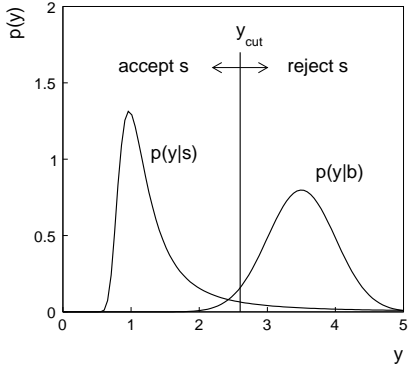}}
\put(9.0,.8){\makebox(6,4)[b]{\begin{minipage}[b]{6cm}
\protect\caption{{\small Distributions of the scalar
test statistic $y(\boldvec{x})$ under the signal and background
hypotheses.}
\protect\label{fig:TestStat}}
\end{minipage}}}
\end{picture}
\end{figure}
\renewcommand{\baselinestretch}{1}
\small\normalsize

To quantify how good the event selection is, we can define the {\it
efficiency} with which one selects events of a given type as the
probability that an event will fall in the acceptance region.  That
is, the signal and background efficiencies are

\begin{eqnarray}
\label{eq:sigeff}
\varepsilon_{\rm s} & = & P( \mbox{accept event} | \mbox{s} ) 
= \int_{\rm A} f(\boldvec{x} | \mbox{s} ) \, d \boldvec{x} 
= \int_{-\infty}^{y_{\rm cut}} p(y | \mbox{s}) \, dy\;, \\*[0.3 cm]
\varepsilon_{\rm b} & = & P( \mbox{accept event} | \mbox{b} ) 
= \int_{\rm A} f(\boldvec{x} | \mbox{b} ) \, d \boldvec{x} 
= \int_{-\infty}^{y_{\rm cut}} p(y | \mbox{b}) \, dy \;, 
\end{eqnarray}

\noindent where the region of integration A represents the
acceptance region.  

Dividing the space of input variables into two regions where one
accepts or rejects the signal hypothesis is essentially the language
of a frequentist statistical test.  If we regard background as the
`null hypothesis', then the background efficiency is the same as
what in a statistical context would be called the significance level
of the test, or the rate of `type-I error'.  Viewing the signal
process as the alternative, the signal efficiency is then what a
statistician would call the power of the test; it is the probability
to reject the background hypothesis if in fact the signal hypothesis
is true.  Equivalently, this is one minus the rate of `type-II
error'.

The use of a statistical test to distinguish between two classes of
events (signal and background), comes up in different ways.  Sometimes
both event classes are known to exist, and the goal is to select one
class (signal) for further study.  For example, proton--proton
collisions leading to the production of top quarks are a
well-established process.  By selecting these events one can carry out
precise measurements of the top quark's properties such as its mass.
In other cases, the signal process could represent an extension to the
Standard Model, say, supersymmetry, whose existence is not yet
established, and the goal of the analysis is to see if one can do
this.  Rejecting the Standard Model with a sufficiently high
significance level amounts to discovering something new, and of course
one hopes that the newly revealed phenomena will provide important
insights into how Nature behaves.

What the physicist would like to have is a test with maximal power
with respect to a broad class of alternative hypotheses.  For two
specific signal and background hypotheses, it turns out that there is
a well defined optimal solution to our problem.  The {\it
Neyman--Pearson} lemma states that one obtains the maximum power
relative for the signal hypothesis for a given significance level
(background efficiency) by defining the acceptance region such that,
for $\boldvec{x}$ inside the region, the {\it likelihood ratio}, i.e.,
the ratio of pdfs for signal and background,

\begin{equation}
\label{eq:lratio}
\lambda( \boldvec{x} ) = 
\frac{f(\boldvec{x} | \mbox{s} )}{ f(\boldvec{x} | \mbox{b} )} \;,
\end{equation}

\noindent is greater than or equal to a given constant, and it is less
than this constant everywhere outside the acceptance region.  This is
equivalent to the statement that the ratio~(\ref{eq:lratio})
represents the test statistic with which one obtains the highest
signal efficiency for a given background efficiency, or equivalently,
for a given signal purity.

In principle the signal and background theories should allow us to
work out the required functions $f(\boldvec{x} | \mbox{s})$ and
$f(\boldvec{x} | \mbox{b})$, but in practice the calculations are too
difficult and we do not have explicit formulae for these.
What we have instead of $f(\boldvec{x} | \mbox{s})$ and $f(\boldvec{x}
| \mbox{b})$ are complicated Monte Carlo programs, that is, we can
sample $\boldvec{x}$ to produce simulated signal and background
events.  Because of the multivariate nature of the data, where
$\boldvec{x}$ may contain at least several or perhaps even hundreds of
components, it is a nontrivial problem to construct a test with a
power approaching that of the likelihood ratio.

In the usual case where the likelihood ratio~(\ref{eq:lratio}) cannot
be used explicitly, there exists a variety of other multivariate
classifiers that effectively separate different types of events.
Methods often used in HEP include {\it neural networks} or {\it Fisher
discriminants}.  Recently, further classification methods from
machine learning have been applied in HEP analyses; these include {\it
probability density estimation (PDE)} techniques, {\it kernel-based
PDE} ({\it KDE} or {\it Parzen window}), {\it support vector
machines}, and {\it decision trees}.  Techniques such as `boosting'
and `bagging' can be applied to combine a number of classifiers into
a stronger one with greater stability with respect to fluctuations in
the training data.  Descriptions of these methods can be found, for example, in the textbooks~\cite{Bishop,Hastie,Duda,Webb} and in Proceedings
of the PHYSTAT conference series~\cite{PHYSTAT}.  Software for HEP
includes the {\tt TMVA}~\cite{TMVA} and 
{\tt StatPatternRecognition}~\cite{Narsky05} packages, although support for the latter has
unfortunately been discontinued.

As we will not have the time to examine all of the methods mentioned
above, in the following section we look at a specific example of a
classifier to illustrate some of the main ideas of a multivariate
analysis: the boosted decision tree (BDT).

\subsection{Boosted decision trees}
\label{sec:bdt}

Boosted decision trees exploit relatively recent developments in
machine learning and have gained significant popularity in HEP.  First
in Section~\ref{sec:dectree} we describe the basic idea of a decision
tree, and then in Section~\ref{sec:boosting} we will say how the the
technique of `boosting' can be used to improve its performance.

\subsubsection{Decision trees}
\label{sec:dectree}

A decision tree is defined by a collection of successive cuts on the
set of input variables.  To determine the appropriate cuts, one begins
with a sample of $N$ training events which are known to be either
signal or background, e.g., from Monte Carlo.  The set of $n$ input
variables measured for each event constitutes a vector $\boldvec{x} =
(x_1, \ldots x_n)$.  Thus we have $N$ instances of $\boldvec{x}$,
$\boldvec{x}_1, \ldots \boldvec{x}_N$, as well as the corresponding
$N$ true class labels $y_{1}, \ldots, y_{N}$.  It is convenient to
assign numerical values to the labels so that, e.g., $y=1$ corresponds
to signal and $y=-1$ for background.

In addition we will assume that each event can be assigned a weight,
$w_i$, with $i = 1, \ldots, N$.  For any subset of the events and for
a set of weights, the signal fraction (purity) is taken to be

\begin{equation}
\label{eq:dectreepurity}
p = \frac{\sum_{i \in \mbox{s}} w_i }
{\sum_{i \in \mbox{s}} w_i  + \sum_{i \in \mbox{b}} w_i} \;,
\end{equation}

\noindent where s and b refer to the signal and background event
types.  The weights are not strictly speaking necessary for a decision
tree, but will be used in connection with boosting in
Section~\ref{sec:boosting}.  For a decision tree without boosting we
can simply take all the weights to be equal.

To quantify the degree of separation achieved by a classifier for a
selected subset of the events one can use, for example, the {\it Gini
coefficient}~\cite{gini}, which historically has been used as a
measure of dispersion in economics and is defined as

\begin{equation}
\label{eq:gini}
G = p (1 - p) \;.
\end{equation}

\noindent The Gini coefficient is zero if the selected sample is
either pure signal or background.  Another measure is simply the
misclassification rate,

\begin{equation}
\label{eq:misclassrate}
\varepsilon = 1 - \mbox{max}(p, 1-p) \;.
\end{equation}

The idea behind a decision tree is illustrated in
Fig.~\ref{fig:MiniBooneBDT}, from an analysis by the MiniBooNE
neutrino oscillation experiment at Fermilab~\cite{miniboone}.

\setlength{\unitlength}{1.0 cm}
\renewcommand{\baselinestretch}{0.8}
\begin{figure}[htbp]
\begin{picture}(10.0,6)
\put(.5,-2){\includegraphics{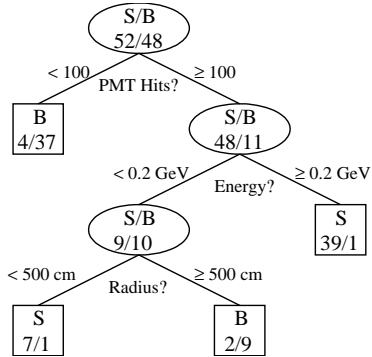}}
\put(9.0,.8){\makebox(6,4)[b]{\begin{minipage}[b]{6cm}
\protect\caption{{\small Illustration of a decision
tree used by the MiniBooNE experiment~\cite{miniboone} (see text).}
\protect\label{fig:MiniBooneBDT}}
\end{minipage}}}
\end{picture}
\end{figure}
\renewcommand{\baselinestretch}{1}
\small\normalsize

One starts with the entire sample of training events in the root node,
shown in the figure with 52 signal and 48 background events.  Out of
all of the possible input variables in the vector $\boldvec{x}$, one
finds the component that provides the best separation between signal
and background by use of a single cut.  This requires a definition of
what constitutes `best separation', and there are a number of
reasonable choices.  For example, for a cut that splits a set of
events $a$ into two subsets $b$ and $c$, one can define the degree of
separation through the weighted change in the Gini coefficients,

\begin{equation}
\label{eq:delta}
\Delta  = W_{a} G_{a} - W_{b} G_{b} - W_{c} G_{c} \;.
\end{equation}

\noindent where

\begin{equation}
W_{a} = \sum_{i \in a} w_i \;,
\end{equation}

\noindent and similarly for $W_{b}$ and $W_{c}$.  Alternatively one
may use a quantity similar to~(\ref{eq:delta}) but with the
misclassification rate~(\ref{eq:misclassrate}), for example, instead of the Gini
coefficient.  More possibilities can be found in Ref.~\cite{TMVA}.

For whatever chosen measure of degree of separation, $\Delta$, one
finds the cut on the variable amongst the components of $\boldvec{x}$
that maximizes it.  In the example of the MiniBooNE experiment shown
in Fig.~\ref{fig:MiniBooneBDT}, this happened to be a cut on the
number of PMT hits with a value of 100.  This splits the training
sample into the two daughter nodes shown in the figure, one of which
is enhanced in signal and the other in background events.

The algorithm requires a stopping rule based, for example, on the number of
events in a node or the misclassification rate.  If, for example, the
number of events or the misclassification rate in a given node falls
below a certain threshold, then this is defined as a terminal node or
`leaf'.  It is classified as a signal or background leaf based on
its predominant event type.  In Fig.~\ref{fig:MiniBooneBDT}, for
example, the node after the cut on PMT hits with 4 signal and 37
background events is classified as a terminal background node.

For nodes that have not yet reached the stopping criterion, one
iterates the procedure and finds, as before, the variable that
provides the best separation with a single cut.  In
Fig.~\ref{fig:MiniBooneBDT} this is an energy cut of $0.2 \,
\mbox{GeV}$.  The steps are continued until all nodes reach the
stopping criterion.

The resulting set of cuts effectively divides the $\boldvec{x}$ space into
two regions: signal and background.  To provide a numerical output
for the classifier we can define

\begin{equation}
\label{eq:ftree}
f(\boldvec{x}) = \left\{ \! \! \begin{array}{ll}
               1 & \quad \boldvec{x} \mbox{ in signal region} , \\*[0.2 cm]
               -1 & \quad \boldvec{x} \mbox{ in background region} .
              \end{array} 
       \right. 
\end{equation}

Equation~(\ref{eq:ftree}) defines a decision tree classifier.  In this
form, these tend to be very sensitive to statistical fluctuations in
the training data.  One can easily see why this is, for example, if
two of the components of $\boldvec{x}$ have similar discriminating
power between signal and background.  For a given training sample, one
variable may be found to give the best degree of separation and is
chosen to make the cut, and this affects the entire further structure
of the tree.  In a different statistically independent sample of
training events, the other variable may be found to be better, and the
resulting tree could look very different.  Boosting is a technique
that can decrease the sensitivity of a classifier to such
fluctuations, and we describe this in the following section.

\subsubsection{Boosting}
\label{sec:boosting}

Boosting is a general method of creating a set of classifiers which
can be combined to give a new classifier that is more stable and has a
smaller misclassification rate than any individual one.  It is often
applied to decision trees, precisely because they suffer from
sensitivity to statistical fluctuations in the training sample, but
the technique can be applied to any classifier.

Let us suppose as above that we have a sample of $N$ training events,
i.e., $N$ instances of the data vector, $\boldvec{x}_1, \ldots,
\boldvec{x}_N$, and $N$ true class labels $y_1, \ldots, y_N$, with
$y=1$ for signal and $y= -1$ for background.  Also as above assume we
have $N$ weights $w_1^{(1)}, \ldots, w_N^{(1)}$, where the superscript
$(1)$ refers to the fact that this is the first training set.  We
initially set the weights equal and normalized such that

\begin{equation}
\label{eq:weightsum}
\sum_{i=1}^N w_i^{(1)} = 1 \;.
\end{equation}

The idea behind boosting is to create from the initial sample, a
series of further training samples which differ from the initial one
in that the weights will be changed according to a specific rule.  A
number of boosting algorithms have been developed, and these differ
primarily in the rule used to update the weights.  We will describe
the AdaBoost algorithm of Freund and Schapire~\cite{freundschapire},
as it was one of the first such algorithms and its properties have
been well studied.

One begins with the initial training sample and from it derives a
classifier.  We have in mind here a decision tree, but it could be any
type of classifier for where the training employs the event weights.
The resulting function $f_1(\boldvec{x})$ will have a certain
misclassification rate $\varepsilon_1$.  In general for the $k$th
classifier (i.e., based on the $k$th training sample), we can write
the error rate as

\begin{equation}
\label{eq:errorrate}
\varepsilon_k = \sum_{i=1}^N w_i^{(k)} I(y_i f_k(\boldvec{x}_i) \le 0) \;,
\end{equation}

\noindent where $I(X) = 1$ if the Boolean expression $X$ is true, and
is zero otherwise.  We then assign a score to the classifier based on
its error rate.  For the AdaBoost algorithm this is 

\begin{equation}
\label{eq:alphak}
\alpha_k = \ln \frac{1 - \varepsilon_k}{\varepsilon_k} \;,
\end{equation}

\noindent which is positive as long as the error rate is lower than
50\%, i.e,. the classifier does better than random guessing.

Having carried out these steps for the initial training sample, we
define the second training sample by updating the weights.  More
generally, the weights for step $k+1$ are found from those for step
$k$ by

\begin{equation}
\label{eq:weightupdate}
w_i^{(k+1)} = w_i^{(k)} \, \frac{ e^{- \alpha_k f_k(\boldvec{x}_i) y_i / 2}}
{Z_k} \;,
\end{equation}

\noindent where the factor $Z_k$ is chosen so that the sum of the
updated weights is equal to unity.  Note that if an event is
incorrectly classified, then the true class label $y_i$ and the value
$f_k(\boldvec{x}_i)$ have opposite signs, and thus the new weights are
greater than the old ones.  Correctly classified events have their
weights decreased.  This means that the updated training set will pay
more attention in the next iteration to those events that were not
correctly classified, the idea being that it should try harder to get
it right the next time around.

After $K$ iterations of this procedure one has classifiers
$f_1(\boldvec{x}), \ldots, f_K(\boldvec{x})$, each with a certain error rate
and score based on Eqs.~(\ref{eq:errorrate}) and~(\ref{eq:alphak}).
In the case of decision trees, the set of new trees is called a {\it
forest}.  From these one defines an averaged classifier as

\begin{equation}
\label{eq:boostave}
y(\boldvec{x}) = \sum_{k=1}^K \alpha_k f_k(\boldvec{x}) \;.
\end{equation}

\noindent Equation~(\ref{eq:boostave}) defines a boosted decision tree
(or more generally, a boosted version of whatever classifier was
used).

One of the important questions to be addressed is how many boosting
iterations to use.  One can show that for a sufficiently large number
of iterations, a boosted decision tree will eventually classify all of
the events in the training sample correctly.  Similar behaviour is
found with any classification method where one can control to an
arbitrary degree the flexibility of the decision boundary.  The user
can arrange it so that the boundary twists and turns so as to get all of
the events on the right side.

In the case of a neural network, for example, one can increase the
number of hidden layers, or the number of nodes in the hidden layers;
for a support vector machine, one can adjust the width of the kernel
function and the regularization parameter to increase the flexibility
of the boundary.  An example is shown in
Fig.~\ref{fig:overTraining}(a), where an extremely flexible classifier
has managed to enclose all of the signal events and exclude all of the
background.

\setlength{\unitlength}{1.0 cm}
\renewcommand{\baselinestretch}{0.9}
\begin{figure}[htbp]
\begin{picture}(10.0,6.5)
\put(0.5,0){\includegraphics{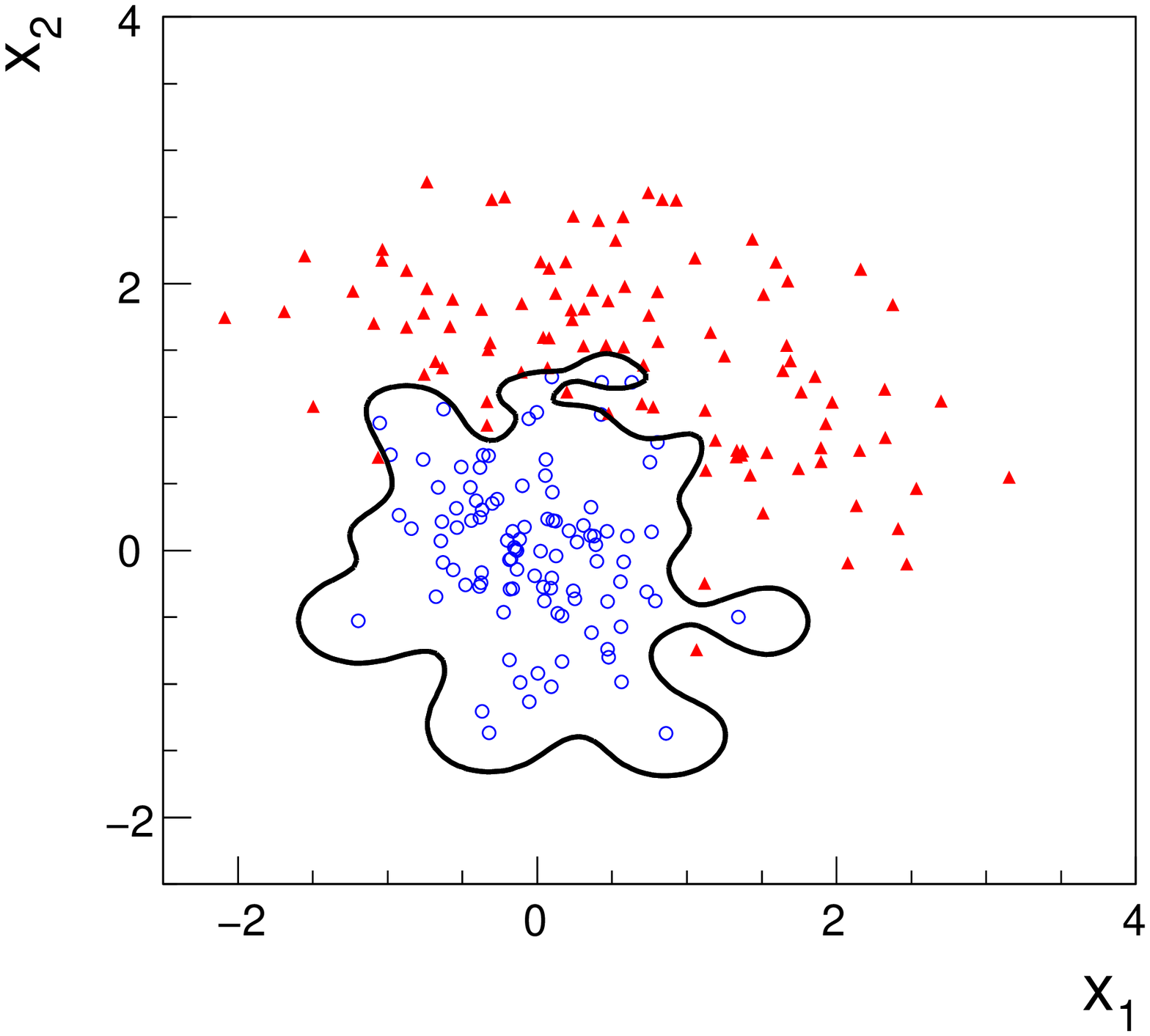}}
\put(8.5,0){\includegraphics{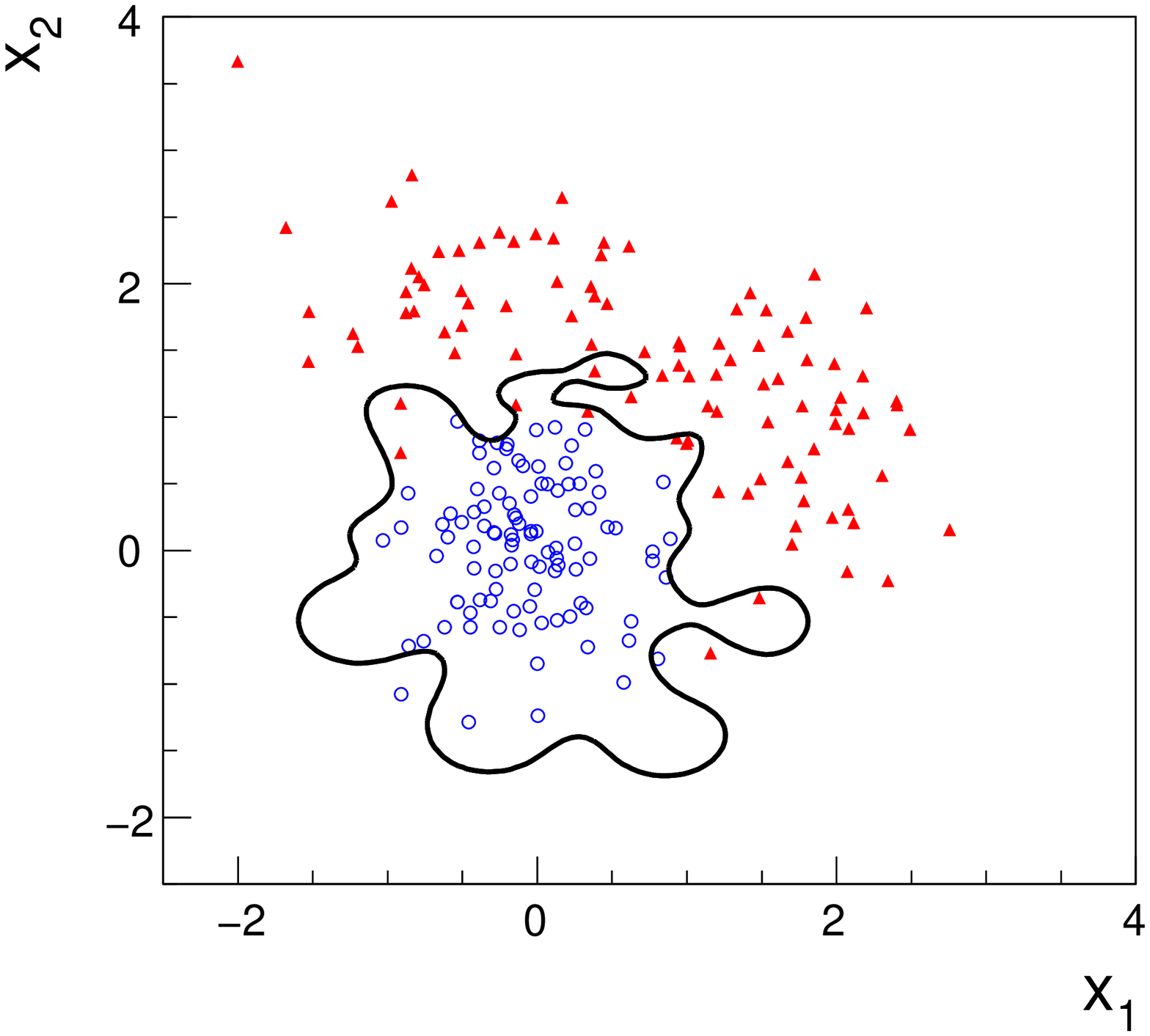}}
\put(0.,5.5){(a)}
\put(15.5,5.5){(b)}
\end{picture}
\caption{\small Scatter plot of events of two types and
the decision boundary determined by a particularly flexible
classifier.  Plot (a) shows the events used to train the
classifier, and (b) shows an independent sample of test data.}
\label{fig:overTraining}
\end{figure}
\renewcommand{\baselinestretch}{1}
\small\normalsize

Of course if we were now to take the decision boundary shown in
Fig.~\ref{fig:overTraining}(a) and apply it to a statistically
independent data sample, there is no reason to believe that the
contortions that led to such good performance on the training sample
will still work.  This can be seen in Fig.~\ref{fig:overTraining}(b),
which shows the same boundary with a new data sample.  In this case
the classifier is said to be {\it overtrained}.  Its error rate
calculated from the same set of events used to train the classifier
underestimates the rate on a statistically independent sample.

To deal with overtraining, one estimates the misclassification rate
not only with the training data sample but also with a statistically
independent test sample.  We can then plot these rates as a function
of the parameters that regulate the flexibility of the decision
boundary, e.g., the number of boosting iterations used to form the
BDT.  For a small number of iterations, one will find in general that
the error rates for both samples drop.  The error rate based on the
training sample will continue to drop, eventually reaching zero.  But
at some point the error rate from the test sample will cease to
decrease and in general will increase.  One chooses the architecture
of the classifier (number of boosting iterations, number of nodes or
layers in a neural network, etc.) to minimize the error rate on the
test sample.

As the test sample is used to choose between a number of competing
architectures based on the minimum observed error rate, this in fact
gives a biased estimate of the true error rate.  In principle one should
use a third validation sample to obtain an unbiased estimate of the
error rate.  In many cases the bias is small and this last step is
omitted, but one should be aware of its potential existence.

In some applications, the training data is relatively inexpensive; one
simply generates more events with Monte Carlo.  But often event
generation can take a prohibitively long time and one may be reluctant
to use only a fraction of the events for training and the other half
for testing.  In such cases, procedures such as {\it cross validation}
(see, e.g., Refs.~\cite{Bishop,Hastie}) can be used where the available
events are partitioned in a number of different ways into training and
test samples and the results averaged.

Boosted decision trees have become increasingly popular in particle
physics in recent years.  One of their advantages is that they are
relatively insensitive to the number of input variables used in the
data vector $\boldvec{x}$.  Components that provide little or no
separation between signal and background are rarely chosen as for the
cut that provides separation, i.e., to split the tree, and thus they
are effectively ignored.  Decision trees have no difficulty in dealing
with different types of data; these can be real, integer, or they can
simply be labels for which there is no natural ordering (categorical
data).  Furthermore, boosted decision trees are surprisingly
insensitive to overtraining.  That is, although the error rate on the
test sample will not decrease to zero as one increases the number of
boosting iterations (as is the case for the training sample), it tends
not to increase.  Further discussion of this point can be found in
Ref.~\cite{freundintro}.

\subsection{Summary on multivariate methods}
\label{sec:mvasum}

The boosted decision tree is an example of a relatively modern
development in Machine Learning that has attracted substantial
attention in HEP.  Support Vector Machines (SVMs) represent another
such development and will no doubt also find further application in
particle physics; further discussion on SVMs can be found in
Refs.~\cite{Bishop,Hastie} and references therein.  Linear classifiers and
neural networks will no doubt continue to play an important role, as
will probability density estimation methods used to approximate the
likelihood ratio.

Multivariate methods have the advantage of exploiting as much
information as possible out of all of the quantities measured for each
event.  In an environment of competition between experiments, this can
be a natural motivation to use them.  Some caution should be
exercised, however, before placing too much faith in the performance
of a complicated classifier, to say nothing of a combination of
complicated classifiers.  These may have decision boundaries that
indeed exploit nonlinear features of the training data, often based on
Monte Carlo.  But if these features have never been verified
experimentally, then they may or may not be present in the real data.
There is thus the risk of, say, underestimating the rate of background
events present in a region where one looks for signal, which could
lead to a spurious discovery.  Simpler classifiers are not immune to
such dangers either, but in such cases the problems may be easier to
control and mitigate.

One should therefore keep in mind the following quote, often heard in
the multivariate analysis community:

\begin{quotation}
{\it Keep it simple. As simple as possible.  Not any simpler.}

--- A.~Einstein
\end{quotation}

\noindent To this we can add the more modern variant,

\begin{quotation}
{\it If you believe in something you don't understand, you suffer, \ldots }

---Stevie Wonder
\end{quotation}

Having made the requisite warnings, however, it seems clear that
multivariate methods will play an important role in the discoveries we
hope to make at the LHC.  One can easily imagine, for example, that
5-sigma evidence for New Physics from a highly performant, and
complicated, classifier would be regarded by the community with some
scepticism.  But if this is backed up by, say, 4-sigma significance
from a simpler, more transparent analysis, then the conclusion would
be more easily accepted, and the team that pursues both approaches may
well win the race.

\section{Summary and conclusions}
\label{sec:summary}

In these lectures we have looked at two topics in statistics, Bayesian
methods and multivariate analysis, which will play an important role
in particle physics in the coming years.  Bayesian methods provide
important tools for analysing systematic uncertainties, where prior
information may be available that does not necessarily stem solely
from other measurements, but rather from theoretical arguments or
other indirect means.  The Bayesian framework allows one to
investigate how the posterior probabilities change upon variation of
the prior probabilities.  Through this type of sensitivity analysis, a
Bayesian result becomes valuable to the broader scientific community.

As experiments become more expensive and the competition more intense,
one will always be looking for ways to exploit as much information as
possible from the data.  Multivariate methods provide a means to
achieve this, and advanced tools such as boosted decision trees have
in recent years become widely used.  And while their use will no doubt
increase as the LHC experiments mature, one should keep in mind that a
simple analysis also has its advantages.  As one studies the advanced
multivariate techniques, however, their properties become more
apparent and the community will surely find ways of using them so as
to maximize the benefits without excessive risk.

\section*{Acknowledgements}

I wish to convey my thanks to the students and organizers of the
2009 European School of High-Energy Physics in Bautzen for a highly stimulating
environment.  The friendly atmosphere and lively discussions created a
truly enjoyable and productive school.

\end{document}